\documentclass[aps,pra,preprint,onecolumn,superscriptaddress,longbibliography,floatfix]{revtex4-2}

\usepackage{amsmath}
\usepackage{amsthm}
\usepackage{amsfonts}
\usepackage{amssymb}
\usepackage{xcolor,graphicx}
\usepackage{tikz}
\usepackage{color}
\usepackage[pdftex]{hyperref}
\usepackage{longtable}
\usepackage{array}
\usepackage{dsfont}
\usepackage{bm}
\usepackage{etoolbox}

\begin{document}

\title{Microscopic exceptional points in the post-selected open Jaynes--Cummings model}

    \author{{Blas Manuel} {Rodr\'iguez-Lara}}
    \email[e-mail: ]{blas.rodriguez@gmail.com}
    \affiliation{Universidad Polit\'ecnica Metropolitana de Hidalgo, Tolcayuca, Hidalgo 43860, Mexico.}

    \date{\today}

\begin{abstract}

Phenomenological non-Hermitian Hamiltonians track selected signatures of complex reservoir dynamics, while post-selected no-jump effective Hamiltonians derived from microscopic open-system theory reveal the underlying system--reservoir physics.
We derive the conditioned Jaynes--Cummings spectrum and propagator at finite temperature using Moore--Penrose normalized generators on the full Hilbert space.
Within the Born--Markov approximation, full Bohr-frequency resolution excludes exceptional points; partial resolution permits coalescence when transition splittings compete with relaxation and the reservoir responses vary negligibly over the coupled transitions.
For zero-temperature examples on shifted resonance, where the bare detuning compensates the reservoir-induced differential Lamb shift, unequal independent losses produce an excitation-dependent threshold, while equal shared losses with vanishing cross Lamb shifts and an orthogonal coupling phase produce a threshold independent of excitation number within the partial-resolution regime.
Our shared-loss examples show oscillatory conditional transfer below the common threshold, monotonic transfer toward unit population at it, and a lower asymptotic population above it.

    \end{abstract}
      
    \maketitle
    \newpage

\section{Introduction}
\label{sec:Sec1}

Exceptional points are spectral singularities of linear operators~\cite{Kato1966}, where eigenvalues and their associated eigenvectors coalesce simultaneously, making the operator defective rather than merely degenerate~\cite{Heiss1990p1167,Heiss2012p444016}.
They have been observed in classical wave platforms, including microwave cavities~\cite{Dembowski2001p787}, optical waveguides~\cite{Guo2009p093902,Ruter2010p192}, and whispering-gallery microresonators~\cite{Peng2014p394}, and in post-selected quantum platforms, including lossy waveguide beamsplitters~\cite{QuirozJuarez2019p862} and superconducting circuits~\cite{Naghiloo2019p1232}.
Near an exceptional point of order $n$, the eigenvalue response to a generic small perturbation $\varepsilon$ scales as $\varepsilon^{1/n}$~\cite{Wiersig2014p203901}.
This spectral response has motivated exceptional point sensing, with demonstrations in optical microcavities~\cite{Chen2017p192,Hodaei2017p187}.

The Jaynes--Cummings model~\cite{Jaynes1963p89} is the canonical excitation-conserving description of a single bosonic mode coupled to a two-level system, with realizations in cavity QED~\cite{Thompson1992p1132}, trapped-ion~\cite{Leibfried2003p281}, and superconducting circuit~\cite{Wallraff2004p162,Blais2021p025005} experiments.
For its open-system dynamics, there exist microscopic master equations for cavity loss only~\cite{Scala2007p013811,Gonzalez2018p015301}, including a finite-temperature equation using a global operator approach~\cite{OzakiNakazato2021}, and quantum-jump and quantum-trajectory descriptions~\cite{Dalibard1992p580,Carmichael1993p2273}.
Under ideal monitoring of emission and absorption channels, post-selection on trajectories with no detected jumps removes the recycling part of the master-equation generator and conditions evolution under a non-Hermitian Hamiltonian, for example, in the Jaynes--Cummings model with cavity loss~\cite{Plenio1998p101}.

The spectrum of the non-Hermitian Hamiltonian for conditioned open-system dynamics defines Hamiltonian exceptional points, while the full Lindblad generator, including quantum jumps, defines Liouvillian exceptional points~\cite{Minganti2019p062131}.
Hybrid Liouvillians describe post-selection with finite detection efficiency~\cite{Minganti2020p062112}, and unobserved quantum jumps enhance decoherence near the exceptional point of a dissipative superconducting qubit~\cite{Chen2021p140504}.
A phenomenological non-Hermitian model with independent cavity and two-level-system losses reproduces the single-excitation exceptional point observed in post-selected qubit--resonator experiments~\cite{Han2025p2400446}.
When the cavity and the two-level system couple to the same reservoir modes, interference between their decay channels can enhance or suppress the decay of the dressed states~\cite{Wang2014p013809}.

Here, we examine how reservoir correlations change the conditioned spectrum of a Jaynes--Cummings system coupled to a common bosonic reservoir, starting from a microscopic derivation of dissipation at finite temperature.
We first introduce in Sec.~\ref{sec:Sec2} a Moore--Penrose normalized \(\mathrm{su}(2)\) representation that diagonalizes the Jaynes--Cummings Hamiltonian on the full Hilbert space with a single operator rotation.
Using our spectral resolution, we derive in Sec.~\ref{sec:Sec3} a Gorini--Kossakowski--Sudarshan--Lindblad master equation.
Full Bohr-frequency resolution excludes exceptional points in our conditioned Hamiltonian.
In Sec.~\ref{sec:Sec4}, we retain interference between unresolved transitions to construct our partially resolved conditioned Hamiltonian.
We derive the finite-temperature conditioned spectrum, exceptional-point condition, and propagator in Sec.~\ref{sec:Sec5} and study the zero-temperature example in Sec.~\ref{sec:Sec6}.
We determine when shared losses produce a common exceptional-point threshold independent of excitation number and examine conditional population transfer below, at, and above this threshold, its post-selection cost, and conjectural connections to experimental platforms.
We close with a summary of our results in Sec.~\ref{sec:Sec7}.

\section{Moore--Penrose Diagonalization}
\label{sec:Sec2}

We start from the Jaynes--Cummings Hamiltonian~\cite{Jaynes1963p89},
\begin{align}
    \hat{H}_{\mathrm{JC}} =&~ \hbar \left[ \omega \hat{a}^{\dagger}\hat{a} + \frac{1}{2} \omega_{0} \hat{\sigma}_{z} + g \left( \hat{a}\hat{\sigma}_{+} + \hat{a}^{\dagger}\hat{\sigma}_{-} \right) \right],
\end{align}
describing the interaction between a bosonic mode of positive frequency $\omega$ with creation (annihilation) operator $\hat{a}^{\dagger}$ ($\hat{a}$) and a two-level system of positive transition frequency $\omega_{0}$ with inversion operator $\hat{\sigma}_{z}$ via the real coupling strength $g$ and ladder operators $\hat{\sigma}_{\pm}$.

The model conserves the total excitation operator,
\begin{align}
    \hat{N} =&~ \hat{a}^{\dagger}\hat{a} + \frac{1}{2}\left( \hat{I} + \hat{\sigma}_{z} \right),
\end{align}
with identity operator $\hat{I}$; our excitation operator commutes with the Hamiltonian, $ \left[ \hat{N}, \hat{H}_{\mathrm{JC}} \right] = 0$, separating the Hilbert space into the zero-excitation sector and the nonzero excitation doublets,
\begin{align}
    \begin{aligned}
        \mathcal{H}_{0} =&~ \mathrm{span}\left\{ \lvert 0,g \rangle \right\}, \\
        \mathcal{H}_{N} =&~ \mathrm{span}\left\{ \lvert N - 1,e \rangle,\lvert N,g \rangle \right\},
    \end{aligned}
\end{align}
where the first entry in each ket is the boson number and the second is the two-level-system state, ground $\lvert g \rangle$ or excited $\lvert e \rangle$, with excitation number $N \in \mathbb{Z}_{+}$ and projectors,
\begin{align}
    \begin{aligned}
        \hat{P}_{0} =&~ \lvert 0, g \rangle \langle 0, g \rvert, \\
        \hat{P}_{N} =&~ \lvert N - 1, e \rangle \langle N - 1, e \rvert + \lvert N, g \rangle \langle N, g \rvert.
    \end{aligned}
\end{align}

We introduce the normalized Jaynes--Cummings generators,
\begin{align}
    \begin{aligned}
        \hat{S}_{x} =&~ \frac{1}{2} \hat{N}_{\mathrm{P}}^{-\frac{1}{2}}\left( \hat{a}\hat{\sigma}_{+} + \hat{a}^{\dagger}\hat{\sigma}_{-} \right), \\
        \hat{S}_{y} =&~ -\frac{i}{2} \hat{N}_{\mathrm{P}}^{-\frac{1}{2}}\left( \hat{a}\hat{\sigma}_{+} - \hat{a}^{\dagger}\hat{\sigma}_{-} \right), \\
        \hat{S}_{z} =&~ \frac{1}{2}\left( \hat{\sigma}_{z} + \hat{P}_{0} \right),
    \end{aligned}
\end{align}
with Moore--Penrose inverse square root~\cite{Moore1920p394,Penrose1955p406,Baksalary2021p9},
\begin{align}
    \hat{N}_{\mathrm{P}}^{-\frac{1}{2}} =&~ \sum_{N = 1}^{\infty}\frac{1}{\sqrt{N}} \hat{P}_{N},
\end{align}
removing the $\sqrt{N}$ dependence from the matrix elements of our generators, which annihilate the zero-excitation state, $\hat{S}_{j}\hat{P}_{0} = 0$.
Our generators preserve total excitation number, $\left[ \hat{N},\hat{S}_{j} \right] = 0$, and close the normalized $\mathrm{su}(2)$ algebra,
\begin{align}
    \left[ \hat{S}_{j},\hat{S}_{k} \right] =&~ i\varepsilon_{jkl} \hat{S}_{l},
\end{align}
with Cartesian indices $j,k,l \in \left\{ x,y,z \right\}$ and Levi--Civita symbol $\varepsilon_{jkl}$.
Our generators anticommute for $j \ne k$, with each square equal to one quarter of the identity on the nonzero-excitation subspace,
\begin{align}
    \left\{ \hat{S}_{j},\hat{S}_{k} \right\} =&~ \frac{1}{2}\delta_{jk}\left( \hat{I} - \hat{P}_{0} \right),
\end{align}
with Kronecker delta $\delta_{jk}$, providing the Casimir invariant,
\begin{align}
    \hat{\bm{S}}^{2} =&~ \hat{S}_{x}^{2} + \hat{S}_{y}^{2} + \hat{S}_{z}^{2} = \frac{3}{4} \left( \hat{I} - \hat{P}_{0} \right),
\end{align}
that commutes with our generators, $\left[ \hat{\bm{S}}^{2},\hat{S}_{j} \right] = 0$, and has eigenvalue $0$ on the zero-excitation sector and $3/4$ on every nonzero excitation doublet, where our generators form a spin-$1/2$ representation.

In our representation, we diagonalize the Jaynes--Cummings Hamiltonian,
\begin{align}
    \hat{H}_{\mathrm{JC}} =&~ \hbar \left[ \omega\left( \hat{N} - \frac{1}{2} \right) + \Delta \left( \hat{S}_{z} - \frac{1}{2} \hat{P}_{0} \right) + 2g \hat{N}^{\frac{1}{2}}\hat{S}_{x} \right],
\end{align}
with detuning $\Delta = \omega_{0} - \omega$ and phase convention $g > 0$, through the unitary change of reference frame,
\begin{align}
    \lvert \psi(t) \rangle =&~ e^{-i\left[ \omega\left( \hat{N} - \frac{1}{2} \right) - \frac{1}{2}\Delta \hat{P}_{0} \right]t} \hat{R}_{y}(\hat{\theta}) \lvert \psi_{\mathrm{D}}(t) \rangle,
\end{align}
where the first factor provides the phase associated with the mean energy at fixed total excitation number, including the zero-excitation energy, and the second has the standard spin-$1/2$ $\mathrm{SU}(2)$ rotation form,
\begin{align}
    \begin{aligned}
        \hat{R}_{y}(\hat{\theta}) =&~ e^{-i \hat{\theta}\hat{S}_{y}}, \\
        =&~ \hat{P}_{0} + \cos \frac{1}{2} \hat{\theta}\left( \hat{I} - \hat{P}_{0} \right) - 2 i \sin \frac{1}{2} \hat{\theta}\hat{S}_{y},
    \end{aligned}
\end{align}
whose operator angle,
\begin{align}
    \begin{aligned}
        \cos \hat{\theta} =&~ \hat{P}_{0} + \Delta \hat{\Omega}_{\mathrm{P}}^{-1}, \\
        \sin \hat{\theta} =&~ 2g \hat{N}^{\frac{1}{2}}\hat{\Omega}_{\mathrm{P}}^{-1},
    \end{aligned}
\end{align}
depends on the nonnegative Rabi frequency operator and its Moore--Penrose inverse,
\begin{align}
    \begin{aligned}
        \hat{\Omega} =&~ \left[ \Delta^{2}\left( \hat{I} - \hat{P}_{0} \right) + 4 g^{2} \hat{N} \right]^{\frac{1}{2}}, \\
        \hat{\Omega}_{\mathrm{P}}^{-1} =&~ \sum_{N = 1}^{\infty} \left( \Delta^{2} + 4 g^{2} N \right)^{-\frac{1}{2}} \hat{P}_{N},
    \end{aligned}
\end{align}
where our principal operator angle commutes with our generators, $\left[ \hat{\theta},\hat{S}_{j} \right] = 0$, and has eigenvalue $0$ on the zero-excitation sector and an eigenvalue in $\left( 0, \pi \right)$ on every nonzero excitation doublet.
At resonance, our rotation mixes the two bare states with equal weights and the dressed-state energy splitting is $2\hbar g\sqrt{N}$.
The effective Schr\"odinger equation in this frame,
\begin{align}
    i\hbar\frac{\mathrm{d}}{\mathrm{d}t}\lvert \psi_{\mathrm{D}}(t) \rangle =&~ \hat{H}_{\mathrm{D}}\lvert \psi_{\mathrm{D}}(t) \rangle,
\end{align}
has diagonal effective Hamiltonian,
\begin{align}
    \hat{H}_{\mathrm{D}} =&~ \hbar \hat{\Omega} \hat{S}_{z},
\end{align}
for the full Hilbert space.

We use the diagonal form and our rotation to identify the dressed states in the original frame through the dressed population operator, 
\begin{align}
    \begin{aligned}
        \hat{\Sigma} =&~ \hat{R}_{y}(\hat{\theta}) \hat{S}_{z}\hat{R}_{y}^{\dagger}(\hat{\theta}), \\
        =&~ \left( \Delta \hat{S}_{z} + 2 g \hat{N}^{\frac{1}{2}} \hat{S}_{x} \right) \hat{\Omega}_{\mathrm{P}}^{-1},
    \end{aligned}
\end{align}
with eigenvalue $0$ on the zero-excitation sector and eigenvalues $\pm 1/2$ on every nonzero excitation doublet; its expectation value is half the population difference between the upper and lower dressed branches.
We use its square,
\begin{align}
    \hat{\Sigma}^{2} =&~ \frac{1}{4}\left( \hat{I} - \hat{P}_{0} \right),
\end{align}
to construct the spectral projectors onto these branches,
\begin{align}
    \hat{\mathcal{P}}_{\pm} =&~ 2 \hat{\Sigma}^{2} \pm \hat{\Sigma},
\end{align}
which exclude the zero-excitation state and resolve the identity alongside the vacuum projector $\hat{P}_{0}$, 
\begin{align}
    \hat{\mathcal{P}}_{+} + \hat{\mathcal{P}}_{-} + \hat{P}_{0} =&~ \hat{I}.
\end{align}
Our branch projectors commute with the excitation projectors, $\left[ \hat{P}_{N}, \hat{\mathcal{P}}_{\pm} \right] = 0$, so the products $\hat{P}_{N}\hat{\mathcal{P}}_{\pm}$ select individual dressed energy eigenstates and provide the spectral resolution of the Jaynes--Cummings Hamiltonian,
\begin{align}
    \hat{H}_{\mathrm{JC}} =&~ \hbar \left[ \omega \left( \hat{N} - \frac{1}{2} \right) - \frac{1}{2} \Delta \hat{P}_{0} + \frac{1}{2} \hat{\Omega} \left( \hat{\mathcal{P}}_{+} - \hat{\mathcal{P}}_{-} \right) \right],
\end{align}
where we recover the standard eigenvalues of sector-by-sector diagonalization,
\begin{align}
    \begin{aligned}
        E_{0} =&~ -\frac{1}{2} \hbar \omega_{0}, \\
        E_{N,\pm} =&~ \hbar\omega\left( N - \frac{1}{2} \right) \pm \frac{\hbar}{2} \sqrt{\Delta^{2} + 4 g^{2} N},
    \end{aligned}
\end{align}
from our full-Hilbert space formalism.

\section{Microscopic master equation}
\label{sec:Sec3}

\begin{figure}[t]
\centering
\includegraphics[width=0.75\columnwidth]{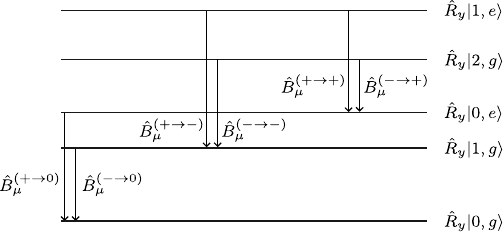}
\caption{
Jaynes--Cummings transition ladder for the first excitation sectors.
The levels show the eigenstates $\hat{R}_{y}(\hat{\theta})\lvert n,s \rangle$, including the vacuum state.
The arrows show the six emission branches labeled by our branch operators $\hat{B}_{\mu}^{(\alpha\rightarrow\beta)}$ with $\mu \in \{ a,\sigma \}$.
The level spacings use $\omega = 1.1\omega_{0}$ and $g = 0.2\omega_{0}$.
}
\label{fig:Fig1}
\end{figure}

Our spectral resolution of the Jaynes--Cummings Hamiltonian,
\begin{align}
    \hat{H}_{\mathrm{JC}} =&~ E_{0} \hat{P}_{0} + \sum_{N = 1}^{\infty} \sum_{\alpha = \pm} E_{N,\alpha} \hat{P}_{N} \hat{\mathcal{P}}_{\alpha},
\end{align}
provides the energy projectors for resolving reservoir-induced transitions.
We couple this system to a bosonic reservoir with Hamiltonian,
\begin{align}
    \hat{H}_{\mathrm{B}} =&~ \hbar \sum_{k} \nu_{k} \hat{b}_{k}^{\dagger} \hat{b}_{k},
\end{align}
with positive frequency $\nu_{k}$ and creation (annihilation) operator $\hat{b}_{k}^{\dagger}$ ($\hat{b}_{k}$) for reservoir mode $k$.
The weak-coupling rotating-wave interaction,
\begin{align}
    \begin{aligned}
        \hat{H}_{\mathrm{JCB}} =&~ \hbar \sum_{k} \left[ \left( \eta_{k}^{(a)} \hat{a} + \eta_{k}^{(\sigma)} \hat{\sigma}_{-} \right) \hat{b}_{k}^{\dagger} + \left( \eta_{k}^{(a)\ast} \hat{a}^{\dagger} + \eta_{k}^{(\sigma)\ast} \hat{\sigma}_{+} \right) \hat{b}_{k} \right],
    \end{aligned}
\end{align}
couples each reservoir mode to the bosonic mode through the complex amplitude $\eta_{k}^{(a)}$ and to the two-level system through $\eta_{k}^{(\sigma)}$, allowing cross correlations between the two reservoir coupling channels~\cite{Wang2014p013809}.
\endgraf
We take the Jaynes--Cummings system initially uncorrelated with a reservoir in thermal equilibrium at finite temperature, $T \geq 0$, with initial joint density operator,
\begin{align}
    \hat{\rho}(0) =&~ \hat{\rho}_{\mathrm{JC}}(0)\otimes \hat{\rho}_{\mathrm{B}},
\end{align}
with initial Jaynes--Cummings density operator $\hat{\rho}_{\mathrm{JC}}(0)$ and thermal reservoir density operator $\hat{\rho}_{\mathrm{B}}$,
\begin{align}
    \begin{aligned}
        \hat{\rho}_{\mathrm{B}} =&~ \frac{1}{Z_{\mathrm{B}}}e^{-\frac{1}{k_{\mathrm{B}}T} \hat{H}_{\mathrm{B}}}, \\
        Z_{\mathrm{B}} =&~ \operatorname{Tr}_{\mathrm{B}}\left[ e^{-\frac{1}{k_{\mathrm{B}}T} \hat{H}_{\mathrm{B}}} \right],
    \end{aligned}
\end{align}
with reservoir partition function $Z_{\mathrm{B}}$, Boltzmann constant $k_{\mathrm{B}}$, and reservoir partial trace $\operatorname{Tr}_{\mathrm{B}}$; the limit $T\rightarrow0^{+}$ provides a zero-temperature reservoir.
\endgraf
We express the dynamics of our complete Hamiltonian,
\begin{align}
    \hat{H} =&~ \hat{H}_{\mathrm{JC}} + \hat{H}_{\mathrm{B}} + \hat{H}_{\mathrm{JCB}},
\end{align}
in the interaction frame, starting from the laboratory frame,
\begin{align}
    \lvert \psi(t) \rangle =&~ e^{-\frac{i}{\hbar} \left( \hat{H}_{\mathrm{JC}} + \hat{H}_{\mathrm{B}} \right) t}\lvert \psi_{\mathrm{I}}(t) \rangle.
\end{align}
The effective interaction Hamiltonian,
\begin{align}
    \hat{H}_{\mathrm{I}}(t) =&~ \hat{H}_{\mathrm{I},+}(t) + \hat{H}_{\mathrm{I},-}(t),
\end{align}
separates into emission and absorption parts,
\begin{align}
    \begin{aligned}
        \hat{H}_{\mathrm{I},+}(t) =&~ \hbar \sum_{k}\sum_{\mu \in \left\{ a,\sigma \right\}} \eta_{k}^{(\mu)}e^{i\nu_{k}t} \hat{A}_{\mu}(t) \hat{b}_{k}^{\dagger}, \\
        \hat{H}_{\mathrm{I},-}(t) =&~ \hat{H}_{\mathrm{I},+}^{\dagger}(t),
    \end{aligned}
\end{align}
where the channel index $\mu \in \{ a,\sigma \}$ labels the bosonic-mode and two-level-system reservoir coupling channels, with channel operators,
\begin{align}
    \hat{A}_{\mu}(t) =&~ e^{\frac{i}{\hbar} \hat{H}_{\mathrm{JC}} t} \hat{A}_{\mu}e^{-\frac{i}{\hbar} \hat{H}_{\mathrm{JC}} t},
\end{align}
with bosonic annihilation operator $\hat{A}_{a} = \hat{a}$ and two-level-system lowering operator $\hat{A}_{\sigma} = \hat{\sigma}_{-}$,
whose excitation-lowering and raising commutators,
\begin{align}
    \begin{aligned}
        \left[ \hat{N},\hat{A}_{\mu} \right] =&~ -\hat{A}_{\mu}, \\
        \left[ \hat{N},\hat{A}_{\mu}^{\dagger} \right] =&~ \hat{A}_{\mu}^{\dagger},
    \end{aligned}
\end{align}
restrict their spectral decomposition to adjacent excitation sectors.
\endgraf
We separate each excitation-lowering operator into the global dressed branches,
\begin{align}
    \hat{B}_{\mu}^{(\alpha\rightarrow\beta)} =&~ \hat{\mathcal{P}}_{\beta}\hat{A}_{\mu}\hat{\mathcal{P}}_{\alpha},
\end{align}
with source index $\alpha = \pm$, target index $\beta \in \{ 0,\pm \}$, and vacuum projector $\hat{\mathcal{P}}_{0} = \hat{P}_{0}$.
The two branches with $\beta = 0$ connect the first excitation doublet to the zero-excitation state $\lvert 0,g \rangle$, the Jaynes--Cummings vacuum, and the four branches with $\beta = \pm$ connect adjacent excitation doublets.
The branch-frequency commutators for emission from $\alpha$ to $\beta$ and absorption from $\beta$ to $\alpha$ through our adjoint branches,
\begin{align}
    \begin{aligned}
        \left[ \hat{H}_{\mathrm{JC}},\hat{B}_{\mu}^{(\alpha\rightarrow\beta)} \right] =&~ -\hbar \hat{\varpi}_{\alpha\beta}\hat{B}_{\mu}^{(\alpha\rightarrow\beta)}, \\
        \left[ \hat{H}_{\mathrm{JC}},\left[ \hat{B}_{\mu}^{(\alpha\rightarrow\beta)} \right]^{\dagger} \right] =&~ \hbar \left[ \hat{B}_{\mu}^{(\alpha\rightarrow\beta)} \right]^{\dagger} \hat{\varpi}_{\alpha\beta},
    \end{aligned}
\end{align}
express the energy difference between the source and target states through the transition-frequency operators,
\begin{align}
    \hat{\varpi}_{\alpha\beta} =&~ \omega \hat{I} + m_{\alpha} \hat{\Omega}_{\uparrow} - m_{\beta} \hat{\Omega} + \frac{\Delta}{2} \hat{P}_{0},
\end{align}
where the eigenvalues of the dressed population operator are $m_{0} = 0$ and $m_{\pm} = \pm1/2$, and the shifted Rabi frequency operator,
\begin{align}
    \hat{\Omega}_{\uparrow} =&~ \left[ \Delta^{2} \hat{I} + 4 g^{2}\left( \hat{N} + \hat{I} \right) \right]^{\frac{1}{2}},
\end{align}
has eigenvalue $\sqrt{\Delta^{2} + 4g^{2}(N + 1)}$ on $\mathcal{H}_{N}$, equal to the Rabi frequency of the source doublet $\mathcal{H}_{N + 1}$.
We obtain the explicit time-dependent channel operators,
\begin{align}
    \hat{A}_{\mu}(t) =&~ \sum_{\alpha = \pm}\sum_{\beta = 0,\pm} e^{-i \hat{\varpi}_{\alpha\beta}t} \hat{B}_{\mu}^{(\alpha\rightarrow\beta)},
\end{align}
through our global projectors, including the vacuum endpoints, without sector-by-sector decomposition.
Our six emission branches have frequencies,
\begin{align}
    \begin{aligned}
        \varpi_{1,\pm0} =&~ \omega + \frac{\Delta}{2} \pm \frac{\Omega_{1}}{2}, \\
        \varpi_{N,\pm\pm} =&~ \omega \pm \frac{1}{2}\left( \Omega_{N} - \Omega_{N - 1} \right), \\
        \varpi_{N,\pm\mp} =&~ \omega \pm \frac{1}{2}\left( \Omega_{N} + \Omega_{N - 1} \right), \qquad N \geq 2,
    \end{aligned}
\end{align}
with Rabi frequency $\Omega_{N} = \sqrt{\Delta^{2} + 4g^{2}N}$ for $N \geq 1$.
For transitions between adjacent doublets, the shifts from $\omega$ depend on the difference of Rabi frequencies within a dressed branch and on their sum between branches.
Figure~\ref{fig:Fig1} shows these transitions for the vacuum and first two excitation doublets; the matrix elements of $\hat{A}_{\mu}$ between the source and target dressed states provide the amplitude of each transition.
\endgraf

We write our emission and absorption Hamiltonians,
\begin{align}
    \begin{aligned}
        \hat{H}_{\mathrm{I},+}(t) =&~ \hbar \sum_{\varpi \in \mathcal{W}}\sum_{\mu \in \left\{ a,\sigma \right\}} e^{-i\varpi t} \hat{B}_{\mu}\left( \varpi \right) \hat{\mathcal{B}}_{\mu}^{\dagger}(t), \\
        \hat{H}_{\mathrm{I},-}(t) =&~ \hbar \sum_{\varpi \in \mathcal{W}}\sum_{\mu \in \left\{ a,\sigma \right\}} e^{i\varpi t} \hat{B}_{\mu}^{\dagger}\left( \varpi \right) \hat{\mathcal{B}}_{\mu}(t),
    \end{aligned}
\end{align}
with Bohr-frequency set $\mathcal{W} = \{ \varpi_{1,\pm0},\varpi_{N,\pm\pm},\varpi_{N,\pm\mp} \}$ for $N \geq 2$ and frequency-resolved channel operators,
\begin{align}
    \begin{aligned}
        \hat{B}_{\mu}\left( \varpi \right) =&~ \sum_{\alpha = \pm}\sum_{\beta = 0,\pm} \hat{\mathbf{1}}_{\{ \varpi \}}\left( \hat{\varpi}_{\alpha\beta} \right) \hat{B}_{\mu}^{(\alpha\rightarrow\beta)},
    \end{aligned}
\end{align}
where the spectral projector $\hat{\mathbf{1}}_{\{ \varpi \}}(\hat{\varpi}_{\alpha\beta})$ projects onto eigenvalue $\varpi$ of the transition-frequency operator.
Each channel operator collects all transitions at that frequency and obeys the commutation relations,
\begin{align}
    \begin{aligned}
        \left[ \hat{H}_{\mathrm{JC}},\hat{B}_{\mu}\left( \varpi \right) \right] =&~ -\hbar\varpi \hat{B}_{\mu}\left( \varpi \right), \\
        \left[ \hat{H}_{\mathrm{JC}},\hat{B}_{\mu}^{\dagger}\left( \varpi \right) \right] =&~ \hbar\varpi \hat{B}_{\mu}^{\dagger}\left( \varpi \right).
    \end{aligned}
\end{align}
The collective reservoir operators,
\begin{align}
    \begin{aligned}
        \hat{\mathcal{B}}_{\mu}^{\dagger}(t) =&~ \sum_{k}\eta_{k}^{(\mu)}e^{i\nu_{k}t} \hat{b}_{k}^{\dagger}, \\
        \hat{\mathcal{B}}_{\mu}(t) =&~ \sum_{k}\eta_{k}^{(\mu)\ast}e^{-i\nu_{k}t} \hat{b}_{k},
    \end{aligned}
\end{align}
create and annihilate excitations in the shared reservoir modes.
For positive Bohr frequencies $\varpi > 0$, emission lowers the system energy by $\hbar\varpi$ and creates a reservoir excitation, while absorption reverses this exchange.
\endgraf
The emission and absorption reservoir correlations~\cite{RivasHuelga2012},
\begin{align}
    \begin{aligned}
        C_{\mu\nu}^{(+)}(\tau) =&~ \operatorname{Tr}_{\mathrm{B}}\left[ \hat{\mathcal{B}}_{\mu}(\tau) \hat{\mathcal{B}}_{\nu}^{\dagger}(0) \hat{\rho}_{\mathrm{B}} \right] \\
        =&~ \int_{0}^{\infty}\mathrm{d}\nu'~J_{\mu\nu}(\nu')\left[ n_{T}(\nu') + 1 \right]e^{-i\nu'\tau}, \\
        C_{\mu\nu}^{(-)}(\tau) =&~ \operatorname{Tr}_{\mathrm{B}}\left[ \hat{\mathcal{B}}_{\mu}^{\dagger}(\tau) \hat{\mathcal{B}}_{\nu}(0) \hat{\rho}_{\mathrm{B}} \right] \\
        =&~ \int_{0}^{\infty}\mathrm{d}\nu'~J_{\nu\mu}(\nu')n_{T}(\nu')e^{i\nu'\tau},
    \end{aligned}
\end{align}
with channel indices $\mu,\nu \in \{ a,\sigma \}$, time difference $\tau$, and reservoir frequency $\nu^{\prime}$, depend on the matrix-valued spectral density~\cite{Breuer2007},
\begin{align}
    J_{\mu\nu}(\nu') =&~ \sum_{k}\eta_{k}^{(\mu)\ast}\eta_{k}^{(\nu)}\delta(\nu' - \nu_{k}),
\end{align}
with Dirac delta $\delta(x)$.
The diagonal entries collect squared coupling amplitudes, while the off-diagonal entries describe the overlap of both channel couplings to shared reservoir modes, including their relative phases; complex conjugation of these products reverses the channel indices in the absorption correlation.
The mean thermal reservoir occupation,
\begin{align}
    n_{T}(\nu') =&~ \frac{1}{e^{\frac{\hbar\nu'}{k_{\mathrm{B}}T}} - 1},
\end{align}
weights stimulated emission and absorption, while the unit term in the emission weight accounts for spontaneous emission.
\endgraf
The half-sided Fourier transforms~\cite{RivasHuelga2012},
\begin{align}
    \int_{0}^{\infty}\mathrm{d}\tau~e^{\pm i\varpi\tau}C_{\mu\nu}^{(\pm)}(\tau) =&~ \frac{1}{2}\Gamma_{\mu\nu}^{(\pm)}(\varpi) + iS_{\mu\nu}^{(\pm)}(\varpi),
\end{align}
provide the emission and absorption rate coefficients at positive Bohr frequency,
\begin{align}
    \begin{aligned}
        \Gamma_{\mu\nu}^{(+)}(\varpi) =&~ 2\pi\left[ n_{T}(\varpi) + 1 \right]J_{\mu\nu}(\varpi), \\
        \Gamma_{\mu\nu}^{(-)}(\varpi) =&~ 2\pi n_{T}(\varpi)J_{\nu\mu}(\varpi),
    \end{aligned}
\end{align}
and Lamb-shift coefficients~\cite{Breuer2007,RivasHuelga2012},
\begin{align}
    \begin{aligned}
        S_{\mu\nu}^{(+)}(\varpi) =&~ \mathcal{P}\int_{0}^{\infty}\mathrm{d}\nu'~\frac{n_{T}(\nu') + 1}{\varpi - \nu'}J_{\mu\nu}(\nu'), \\
        S_{\mu\nu}^{(-)}(\varpi) =&~ -\mathcal{P}\int_{0}^{\infty}\mathrm{d}\nu'~\frac{n_{T}(\nu')}{\varpi - \nu'}J_{\nu\mu}(\nu'),
    \end{aligned}
\end{align}
through the distribution identity $\int_{0}^{\infty}\mathrm{d}\tau\,e^{ix\tau} = \pi\delta(x) + i\mathcal{P}(1/x)$, with Cauchy principal value $\mathcal{P}$ and Fourier frequency difference $x$.
The delta contribution selects resonant reservoir modes, while the principal-value integrals include the full reservoir spectrum for every Bohr frequency $\varpi \in \mathcal{W}$; the opposite Fourier sign accounts for the minus sign in the absorption Lamb shift.
The rates obey thermal detailed balance~\cite{Breuer2007,Trushechkin2021p062226},
\begin{align}
    \Gamma_{\mu\nu}^{(-)}(\varpi) =&~ e^{-\frac{\hbar\varpi}{k_{\mathrm{B}}T}}\Gamma_{\nu\mu}^{(+)}(\varpi),
\end{align}
and vanish for negative Bohr frequencies $\varpi < 0$ in this rotating-wave reservoir coupling.
We evaluate the zero-frequency coefficients from the correlation transforms when those limits exist.
As $T\rightarrow0$, the emission response approaches the vacuum result, and the absorption rates and shifts vanish.
\endgraf
The Gram form~\cite{HornJohnson2013} of the spectral density,
\begin{align}
    \bm{z}^{\dagger}\bm{J}(\nu')\bm{z} =&~ \sum_{k}\left| \sum_{\mu}\eta_{k}^{(\mu)}z_{\mu} \right|^{2}\delta(\nu' - \nu_{k}) \geq 0,
\end{align}
is nonnegative for every complex channel vector $\bm{z}$ with components $z_{\mu}$, where the spectral-density matrix $\bm{J}$ has entries $J_{\mu\nu}$.
The nonnegative thermal weights preserve this positivity in both Kossakowski matrices~\cite{Gorini1976p821,Breuer2007},
\begin{align}
    \bm{\Gamma}^{(\pm)}\left( \varpi \right) =&~
    \begin{pmatrix}
        \Gamma_{aa}^{(\pm)}\left( \varpi \right) & \Gamma_{a\sigma}^{(\pm)}\left( \varpi \right) \\
        \Gamma_{a\sigma}^{(\pm)\ast}\left( \varpi \right) & \Gamma_{\sigma\sigma}^{(\pm)}\left( \varpi \right)
    \end{pmatrix},
\end{align}
with positivity bounds,
\begin{align}
    \begin{aligned}
        \Gamma_{aa}^{(\pm)}\left( \varpi \right),\Gamma_{\sigma\sigma}^{(\pm)}\left( \varpi \right) &\geq 0, \\
        \left\lvert \Gamma_{a\sigma}^{(\pm)}\left( \varpi \right) \right \rvert^{2} &\leq \Gamma_{aa}^{(\pm)}\left( \varpi \right)\Gamma_{\sigma\sigma}^{(\pm)}\left( \varpi \right),
    \end{aligned}
\end{align}
which we use to parametrize the cross-channel coefficients,
\begin{align}
    \Gamma_{a\sigma}^{(\pm)}\left( \varpi \right) =&~ \chi\left( \varpi \right)e^{\pm i\phi\left( \varpi \right)} \sqrt{\Gamma_{aa}^{(\pm)}\left( \varpi \right)\Gamma_{\sigma\sigma}^{(\pm)}\left( \varpi \right)},
\end{align}
with real correlation magnitude $\chi(\varpi) \in [0,1]$ and relative coupling phase $\phi(\varpi)$ in the convention $g > 0$.
At frequency $\varpi$, the two channels couple to orthogonal combinations of reservoir modes at $\chi = 0$ and to the same combination at $\chi = 1$.
\endgraf
We obtain the fully resolved Gorini--Kossakowski--Sudarshan--Lindblad master equation~\cite{Gorini1976p821,Lindblad1976p119,Breuer2007},
\begin{align}
    \frac{\mathrm{d}}{\mathrm{d}t} \hat{\rho}_{\mathrm{I}} =&~ -\frac{i}{\hbar}\left[ \hat{H}_{\mathrm{LS}},\hat{\rho}_{\mathrm{I}} \right] + \mathcal{D}_{\mathrm{I}} \hat{\rho}_{\mathrm{I}},
\end{align}
for the reduced Jaynes--Cummings density operator $\hat{\rho}_{\mathrm{I}}$ in the interaction frame.
We use the Born--Markov approximation with weak system--reservoir coupling and reservoir correlation functions that have convergent time integrals and decay much faster than the system relaxes~\cite{RivasHuelga2012}; for full resolution, we discard terms oscillating at differences between distinct Bohr frequencies, assuming their periods are much shorter than the system relaxation time~\cite{Trushechkin2021p062226}.
Our Lamb-shift Hamiltonian,
\begin{align}
    \begin{aligned}
        \hat{H}_{\mathrm{LS}} =&~ \hbar \sum_{\varpi \in \mathcal{W}}\sum_{\mu,\nu \in \left\{ a,\sigma \right\}} \left[ S_{\mu\nu}^{(+)}\left( \varpi \right) \hat{B}_{\mu}^{\dagger}\left( \varpi \right) \hat{B}_{\nu}\left( \varpi \right) + S_{\mu\nu}^{(-)}\left( \varpi \right) \hat{B}_{\mu}\left( \varpi \right) \hat{B}_{\nu}^{\dagger}\left( \varpi \right) \right],
    \end{aligned}
\end{align}
describes the reservoir-induced shifts of the Jaynes--Cummings energies.
The dissipator,
\begin{align}
    \mathcal{D}_{\mathrm{I}} \hat{\rho}_{\mathrm{I}} =&~ \mathcal{J}_{\mathrm{I}} \hat{\rho}_{\mathrm{I}} - \frac{1}{2}\left\{ \hat{\Gamma}_{\mathrm{I}},\hat{\rho}_{\mathrm{I}} \right\},
\end{align}
contains the recycling contribution~\cite{Carmichael1993p2273},
\begin{align}
    \begin{aligned}
        \mathcal{J}_{\mathrm{I}} \hat{\rho}_{\mathrm{I}} =&~ \sum_{\varpi \in \mathcal{W}}\sum_{\mu,\nu \in \left\{ a,\sigma \right\}} \left[ \Gamma_{\mu\nu}^{(+)}\left( \varpi \right) \hat{B}_{\nu}\left( \varpi \right) \hat{\rho}_{\mathrm{I}}\hat{B}_{\mu}^{\dagger}\left( \varpi \right) + \Gamma_{\mu\nu}^{(-)}\left( \varpi \right) \hat{B}_{\nu}^{\dagger}\left( \varpi \right) \hat{\rho}_{\mathrm{I}}\hat{B}_{\mu}\left( \varpi \right) \right],
    \end{aligned}
\end{align}
which transfers population between adjacent excitation sectors through emission and absorption jumps, and the no-jump decay operator,
\begin{align}
    \begin{aligned}
        \hat{\Gamma}_{\mathrm{I}} =&~ \sum_{\varpi \in \mathcal{W}}\sum_{\mu,\nu \in \left\{ a,\sigma \right\}} \left[ \Gamma_{\mu\nu}^{(+)}\left( \varpi \right) \hat{B}_{\mu}^{\dagger}\left( \varpi \right) \hat{B}_{\nu}\left( \varpi \right) + \Gamma_{\mu\nu}^{(-)}\left( \varpi \right) \hat{B}_{\mu}\left( \varpi \right) \hat{B}_{\nu}^{\dagger}\left( \varpi \right) \right],
    \end{aligned}
\end{align}
whose expectation value in a normalized system state is the total jump rate.
With $\eta_{k}^{(\sigma)} = 0$ and neglecting Lamb shifts, our fully resolved master equation recovers the microscopic cavity-loss form~\cite{Scala2007p013811,Gonzalez2018p015301}.
We write our master equation in the quantum-jump representation,
\begin{align}
    \frac{\mathrm{d} \hat{\rho}_{\mathrm{I}}}{\mathrm{d}t} =&~ -\frac{i}{\hbar}\left( \hat{H}_{\mathrm{NJ}}\hat{\rho}_{\mathrm{I}} - \hat{\rho}_{\mathrm{I}}\hat{H}_{\mathrm{NJ}}^{\dagger} \right) + \mathcal{J}_{\mathrm{I}} \hat{\rho}_{\mathrm{I}},
\end{align}
with no-jump conditioned Hamiltonian,
\begin{align}
    \hat{H}_{\mathrm{NJ}} =&~ \hat{H}_{\mathrm{LS}} - \frac{i\hbar}{2} \hat{\Gamma}_{\mathrm{I}},
\end{align}
which generates the unnormalized evolution conditioned on detecting no emission or absorption jumps under ideal monitoring.
The commutators of the fully resolved channel products,
\begin{align}
    \begin{aligned}
        \left[ \hat{H}_{\mathrm{JC}},\hat{B}_{\mu}^{\dagger}(\varpi) \hat{B}_{\nu}(\varpi) \right] = \left[ \hat{N},\hat{B}_{\mu}^{\dagger}(\varpi) \hat{B}_{\nu}(\varpi) \right] =&~ 0, \\
        \left[ \hat{H}_{\mathrm{JC}},\hat{B}_{\mu}(\varpi) \hat{B}_{\nu}^{\dagger}(\varpi) \right] = \left[ \hat{N},\hat{B}_{\mu}(\varpi) \hat{B}_{\nu}^{\dagger}(\varpi) \right] =&~ 0,
    \end{aligned}
\end{align}
show that our Lamb-shift and decay operators commute with the Jaynes--Cummings Hamiltonian and excitation operator at finite temperature.
Our conditioned Hamiltonian is diagonal in the global dressed basis and has no exceptional points.

\section{Conditioned spectra}
\label{sec:Sec4}

An exceptional point requires non-diagonal reservoir-induced terms in the global dressed basis.
The common reservoir can generate these terms through emission from the two dressed states at excitation number $N$ to a common state at $N - 1$, or absorption to a common state at $N + 1$, when their transition-frequency separation is comparable to the relaxation linewidths~\cite{Cattaneo2019p113045,Trushechkin2021p062226},
\begin{align}
    \Omega_{N} = \left| \varpi - \varpi' \right| &\lesssim \frac{1}{\tau_{\mathrm{R}}},
\end{align}
corresponding to partial Bohr-frequency resolution for near-resonant Jaynes--Cummings dynamics, $\omega \sim \omega_{0}$.
We estimate the shortest relevant relaxation time~\cite{Cattaneo2019p113045},
\begin{align}
    \tau_{\mathrm{R}} &\sim \min_{j}\frac{1}{\Gamma_{j}} = \frac{1}{\max_{j}\Gamma_{j}},
\end{align}
from the positive emission and absorption transition rates $\Gamma_{j}$, which the thermal rate matrices $\bm{\Gamma}^{(\pm)}(\varpi)$ and channel matrix elements provide.
This is an estimate because population and coherence decay involve different rate combinations.
\endgraf
We require the weak-coupling hierarchy,
\begin{align}
    \max\left\{ g\sqrt{N + 1},|\Delta|,\Gamma_{j} \right\} &\ll \min\left\{ \frac{1}{\tau_{\mathrm{B}}},\omega,\omega_{0} \right\},
\end{align}
for the excitation numbers and decay processes involved in the dynamics, including absorption into adjacent excitation sectors.
The reservoir correlation time $\tau_{\mathrm{B}}$ characterizes the decay of the correlations $C_{\mu\nu}^{(\pm)}(\tau)$ and depends on the reservoir spectral profile and temperature~\cite{RivasHuelga2012}.
Coherent exchange and losses can be comparable within this hierarchy, which preserves the Jaynes--Cummings rotating-wave regime~\cite{Jaynes1963p89,Blais2021p025005}.
We choose the coarse-graining interval,
\begin{align}
    \max\left\{ \tau_{\mathrm{B}},\frac{1}{\omega} \right\} \ll \Delta t_{\mathrm{cg}} &\ll \tau_{\mathrm{R}},
\end{align}
to average the carrier oscillations and require $\left| \varpi - \varpi' \right|\Delta t_{\mathrm{cg}} \ll 1$ for every pair of emission frequencies and every pair of absorption frequencies involved in the dynamics, preserving the relative phases of the dressed transitions.
We evaluate the reservoir responses at the common carrier~\cite{Trushechkin2021p062226},
\begin{align}
    \begin{aligned}
        \Gamma_{\mu\nu}^{(\pm)}(\varpi) &\simeq \Gamma_{\mu\nu}^{(\pm)}(\omega) \equiv \Gamma_{\mu\nu}^{(\pm)}, \\
        S_{\mu\nu}^{(\pm)}(\varpi) &\simeq S_{\mu\nu}^{(\pm)}(\omega) \equiv S_{\mu\nu}^{(\pm)},
    \end{aligned}
\end{align}
when the emission and absorption rates and Lamb shifts each vary negligibly over these transition frequencies, including their thermal occupation factors.
\endgraf
Our global channel identities,
\begin{align}
    \begin{aligned}
        \sum_{\varpi \in \mathcal{W}} \hat{B}_{\mu}(\varpi) =&~ \hat{A}_{\mu}, \\
        \sum_{\varpi \in \mathcal{W}} \hat{B}_{\mu}^{\dagger}(\varpi) =&~ \hat{A}_{\mu}^{\dagger},
    \end{aligned}
\end{align}
reconstruct the emission and absorption operators.
Retaining the cross terms between unresolved emission transitions and between unresolved absorption transitions, we obtain the laboratory-frame master equation,
\begin{align}
    \frac{\mathrm{d} \hat{\rho}_{\mathrm{JC}}}{\mathrm{d}t} =&~ -\frac{i}{\hbar} \left[ \hat{H}_{\mathrm{JC}} + \hat{H}_{\mathrm{LS}}^{(\mathrm{p})},\hat{\rho}_{\mathrm{JC}} \right] + \mathcal{D}^{(\mathrm{p})} \hat{\rho}_{\mathrm{JC}},
\end{align}
for the reduced Jaynes--Cummings density operator $\hat{\rho}_{\mathrm{JC}}$, with Lamb-shift Hamiltonian,
\begin{align}
    \hat{H}_{\mathrm{LS}}^{(\mathrm{p})} =&~ \hbar \sum_{\mu,\nu} \left[ S_{\mu\nu}^{(+)} \hat{A}_{\mu}^{\dagger}\hat{A}_{\nu} + S_{\mu\nu}^{(-)} \hat{A}_{\mu}\hat{A}_{\nu}^{\dagger} \right],
\end{align}
and dissipator,
\begin{align}
    \begin{aligned}
        \mathcal{D}^{(\mathrm{p})} \hat{\rho}_{\mathrm{JC}} =&~ \sum_{\mu,\nu}\Gamma_{\mu\nu}^{(+)} \left[ \hat{A}_{\nu}\hat{\rho}_{\mathrm{JC}}\hat{A}_{\mu}^{\dagger} - \frac{1}{2}\left\{ \hat{A}_{\mu}^{\dagger}\hat{A}_{\nu},\hat{\rho}_{\mathrm{JC}} \right\} \right] \\
        &~ + \sum_{\mu,\nu}\Gamma_{\mu\nu}^{(-)} \left[ \hat{A}_{\nu}^{\dagger}\hat{\rho}_{\mathrm{JC}}\hat{A}_{\mu} - \frac{1}{2}\left\{ \hat{A}_{\mu}\hat{A}_{\nu}^{\dagger},\hat{\rho}_{\mathrm{JC}} \right\} \right],
    \end{aligned}
\end{align}
with the partial-resolution superscript $\mathrm{p}$ and channel indices $\mu,\nu \in \{ a,\sigma \}$.
The two positive rate matrices preserve the Gorini--Kossakowski--Sudarshan--Lindblad form~\cite{Gorini1976p821,Lindblad1976p119}.
\endgraf
The non-Hermitian Schr\"odinger equation,
\begin{align}
    i\hbar\frac{\mathrm{d}}{\mathrm{d}t}\lvert \psi \rangle =&~ \hat{H}_{\mathrm{JC}}^{(\mathrm{NJ})}\lvert \psi \rangle,
\end{align}
governs the unnormalized conditioned state $\lvert \psi \rangle$ under ideal monitoring and post-selection on the absence of emission and absorption jumps~\cite{Dalibard1992p580}.
For an initially normalized pure state, its squared norm is the no-jump probability.
Our conditioned Hamiltonian,
\begin{align}
    \hat{H}_{\mathrm{JC}}^{(\mathrm{NJ})} =&~ \hat{H}_{\mathrm{JC}} + \hat{H}_{\mathrm{res}}^{(\mathrm{NJ})},
\end{align}
contains the reservoir correction,
\begin{align}
    \hat{H}_{\mathrm{res}}^{(\mathrm{NJ})} =&~ \hbar \left[ \omega_{T} \hat{I} + \sum_{\mu,\nu} \left( S_{\mu\nu} - \frac{i}{2}\Gamma_{\mu\nu} \right) \hat{A}_{\mu}^{\dagger}\hat{A}_{\nu} \right],
\end{align}
where the complex thermal-shift frequency,
\begin{align}
    \omega_{T} =&~ S_{aa}^{(-)} + S_{\sigma\sigma}^{(-)} - \frac{i}{2}\left( \Gamma_{aa}^{(-)} + \Gamma_{\sigma\sigma}^{(-)} \right),
\end{align}
collects the identity contributions of the absorption products $\hat{a}\hat{a}^{\dagger} = \hat{a}^{\dagger}\hat{a} + \hat{I}$ and $\hat{\sigma}_{-}\hat{\sigma}_{+} = \hat{I} - \hat{\sigma}_{+}\hat{\sigma}_{-}$.
The effective Lamb-shift and no-jump loss coefficients,
\begin{align}
    \begin{aligned}
        S_{\mu\nu} =&~ S_{\mu\nu}^{(+)} + \left( 1 - 2\delta_{\mu\sigma}\delta_{\nu\sigma} \right) S_{\nu\mu}^{(-)}, \\
        \Gamma_{\mu\nu} =&~ \Gamma_{\mu\nu}^{(+)} + \left( 1 - 2\delta_{\mu\sigma}\delta_{\nu\sigma} \right) \Gamma_{\nu\mu}^{(-)},
    \end{aligned}
\end{align}
obey the conjugation relations $S_{\nu\mu} = S_{\mu\nu}^{\ast}$ and $\Gamma_{\nu\mu} = \Gamma_{\mu\nu}^{\ast}$.
The emission and absorption Kossakowski matrices $\bm{\Gamma}^{(\pm)}$ are positive; our effective matrix $\bm{\Gamma}$ need not be positive by itself because the complete no-jump decay operator includes the scalar absorption contribution.
\endgraf
We separate our conditioned Hamiltonian,
\begin{align}
    \hat{H}_{\mathrm{JC}}^{(\mathrm{NJ})} =&~ \hat{H}_{0}^{(\mathrm{NJ})} + \hat{H}_{\mathrm{S}}^{(\mathrm{NJ})},
\end{align}
into the diagonal contribution,
\begin{align}
    \hat{H}_{0}^{(\mathrm{NJ})} =&~ \hbar \left[ \omega_{0}^{(\mathrm{NJ})} \hat{I} - \frac{1}{2} \Delta^{(\mathrm{NJ})} \hat{P}_{0} + \omega^{(\mathrm{NJ})} \hat{N} \right],
\end{align}
with complex frequency offset, detuning, and excitation number frequency,
\begin{align}
    \begin{aligned}
        \omega_{0}^{(\mathrm{NJ})} =&~ \frac{1}{2} \left( \Delta^{(\mathrm{NJ})} - \omega_{0} \right) + \omega_{T}, \\
        \Delta^{(\mathrm{NJ})} =&~ \Delta + \Delta S - \frac{i}{2}\Delta\Gamma, \\
        \omega^{(\mathrm{NJ})} =&~ \omega + S_{aa} - \frac{i}{2}\Gamma_{aa},
    \end{aligned}
\end{align}
with differential Lamb shift and differential no-jump loss rate,
\begin{align}
    \begin{aligned}
        \Delta S =&~ S_{\sigma\sigma} - S_{aa}, \\
        \Delta\Gamma =&~ \Gamma_{\sigma\sigma} - \Gamma_{aa},
    \end{aligned}
\end{align}
and the normalized $\mathrm{su}(2)$ contribution,
\begin{align}
    \hat{H}_{\mathrm{S}}^{(\mathrm{NJ})} =&~ \hbar \left[ \Delta^{(\mathrm{NJ})} \hat{S}_{z} + \hat{N}^{\frac{1}{2}} \left( g_{\sigma a}^{(\mathrm{NJ})} \hat{S}_{+} + g_{a\sigma}^{(\mathrm{NJ})} \hat{S}_{-} \right) \right],
\end{align}
with exchange coefficients,
\begin{align}
    \begin{aligned}
        g_{\sigma a}^{(\mathrm{NJ})} =&~ g + S_{\sigma a} - \frac{i}{2}\Gamma_{\sigma a}, \\
        g_{a\sigma}^{(\mathrm{NJ})} =&~ g + S_{a\sigma} - \frac{i}{2}\Gamma_{a\sigma},
    \end{aligned}
\end{align}
and normalized ladder operators $\hat{S}_{\pm} = \hat{S}_{x} \pm i \hat{S}_{y}$.
Our diagonal Hamiltonian provides the excitation-dependent phase and decay and commutes with our normalized spin Hamiltonian, which determines state mixing and complex-energy splitting.
The vacuum has conditioned energy $\hbar \omega_{T} + E_{0}$ and no-jump decay rate $\Gamma_{aa}^{(-)} + \Gamma_{\sigma\sigma}^{(-)}$ due to thermal absorption.

We write the projection of our normalized spin Hamiltonian onto a nonzero excitation doublet,
\begin{align}
    \hat{P}_{N}\hat{H}_{\mathrm{S}}^{(\mathrm{NJ})}\hat{P}_{N} =&~ \hbar
    \begin{pmatrix}
        \frac{1}{2}\Delta^{(\mathrm{NJ})} & g_{\sigma a}^{(\mathrm{NJ})} \sqrt{N} \\
        g_{a\sigma}^{(\mathrm{NJ})} \sqrt{N} & -\frac{1}{2}\Delta^{(\mathrm{NJ})}
    \end{pmatrix}_{\mathcal{B}_{N}},
\end{align}
in the ordered basis $\mathcal{B}_{N} = \{ \lvert N - 1,e \rangle,\lvert N,g \rangle \}$.
For independent reservoirs, $S_{\sigma a} = S_{a\sigma} = 0$ and $\Gamma_{\sigma a} = \Gamma_{a\sigma} = 0$, we recover the phenomenological normalized spin Hamiltonian,
\begin{align}
    \hat{H}_{\mathrm{S},N}^{(\mathrm{Phen})} =&~ \hbar
    \begin{pmatrix}
        \frac{1}{2}\Delta^{(\mathrm{Phen})} & g\sqrt{N} \\
        g\sqrt{N} & -\frac{1}{2}\Delta^{(\mathrm{Phen})}
    \end{pmatrix}_{\mathcal{B}_{N}},
\end{align}
with complex detuning $\Delta^{(\mathrm{Phen})} = \Delta + \Delta S - i\Delta\kappa_{\mathrm{eff}}/2$ and decay imbalance $\Delta\kappa_{\mathrm{eff}} = \kappa_{\sigma} - \kappa_{a} = \Delta\Gamma$ between the two-level no-jump coefficient $\kappa_{\sigma}$ and bosonic-mode no-jump coefficient $\kappa_{a}$.
A nonzero loss imbalance $\Delta\Gamma$ produces off-diagonal decay terms in the dressed basis even when the cross-channel coefficients vanish.
Our single-excitation, zero-temperature case describes the dissipative qubit--resonator dimer~\cite{Han2025p2400446}.

\section{Exceptional points}
\label{sec:Sec5}

Our conditioned Hamiltonian $\hat H_{\mathrm{JC}}^{(\mathrm{NJ})}$ admits the time-dependent similarity transformation,
\begin{align}
    \lvert\psi(t)\rangle =&~ e^{-\frac{i}{\hbar}\hat H_0^{(\mathrm{NJ})}t}e^{\hat\xi_-\hat S_-}e^{\hat\xi_+\hat S_+}\lvert\psi_{\mathrm D}^{(\mathrm{NJ})}(t)\rangle,
\end{align}
where the first factor provides the phase and amplitude scaling associated with the real and imaginary parts of our diagonal contribution $\hat H_0^{(\mathrm{NJ})}$, and the two remaining factors depend on the dimensionless operator parameters,
\begin{align}
    \begin{aligned}
        \hat\xi_- =&~ 2g_{a\sigma}^{(\mathrm{NJ})} \hat N^{\frac{1}{2}} ~ \left[\hat{\mathcal D}^{\frac{1}{2}}+\Delta^{(\mathrm{NJ})}\left(\hat I-\hat P_0\right)\right]_{\mathrm P}^{-1}, \\
        \hat\xi_+ =&~ -g_{\sigma a}^{(\mathrm{NJ})} \hat N^{\frac{1}{2}} \hat{\mathcal{D}}^{-\frac{1}{2}}_{\mathrm P},
    \end{aligned}
\end{align}
that annihilate the zero-excitation state, $\hat\xi_\pm\lvert0,g\rangle=0$, making their exponential maps preserve it, $e^{\hat\xi_\pm\hat S_\pm}\lvert0,g\rangle=\lvert0,g\rangle$, with discriminant operator,
\begin{align}
    \begin{aligned}
        \hat{\mathcal D} =&~ \frac{4}{\hbar^2} \hat H_{\mathrm S}^{(\mathrm{NJ}) 2} \\
        =&~ \Delta^{(\mathrm{NJ}) 2} \left(\hat I-\hat P_0\right) + 4 g_{a\sigma}^{(\mathrm{NJ})} g_{\sigma a}^{(\mathrm{NJ})} \hat N,
    \end{aligned}
\end{align}
extending our supersymmetric squared-Hamiltonian construction for the Jaynes--Cummings Hamiltonian~\cite{Kafuri2024pC82}.
Our similarity transformation provides the effective Schr\"odinger equation,
\begin{align}
    i\hbar\frac{\mathrm{d}}{\mathrm{d}t}\lvert\psi_{\mathrm D}^{(\mathrm{NJ})}(t)\rangle =&~ \hat H_{\mathrm D}^{(\mathrm{NJ})} \lvert\psi_{\mathrm D}^{(\mathrm{NJ})}(t)\rangle,
\end{align}
where the time-dependent factor removes the diagonal contribution $\hat H_0^{(\mathrm{NJ})}$ and the effective Hamiltonian takes the form,
\begin{align}
    \hat H_{\mathrm D}^{(\mathrm{NJ})} =&~
    \begin{cases}
        \hbar\hat{\mathcal D}^{\frac{1}{2}}\hat S_z & \hat{\mathcal D}\ne0, \hat{\mathcal D}^{\frac{1}{2}}\ne-\Delta^{(\mathrm{NJ})}, \\
        \hat{H}_{\mathrm{S}}^{(\mathrm{NJ})} + \hbar g_{\sigma a}^{(\mathrm{NJ})}\hat{N}^{\frac{1}{2}}\hat{S}_{+} & \hat{\mathcal D}^{\frac{1}{2}}=-\Delta^{(\mathrm{NJ})}\ne0, \\
        \hbar g_{\sigma a}^{(\mathrm{NJ})}\hat N^{\frac{1}{2}}\hat S_+ & \hat{\mathcal D}=0, \Delta^{(\mathrm{NJ})}\ne0, \\
        \hat H_{\mathrm S}^{(\mathrm{NJ})}, & \hat{\mathcal D}=0, \Delta^{(\mathrm{NJ})}=0.
    \end{cases}
\end{align}
The first case has diagonal form with energy eigenvalues
\begin{align}
    \begin{aligned}
        \lambda_{0} =&~ 0, \\
        \lambda_{N,\pm}^{(\mathrm{NJ})} =&~ \pm\frac{1}{2} \hbar \sqrt{\mathcal D_N},
    \end{aligned}
\end{align}
where the discriminant eigenvalue
\begin{align}
    \mathcal{D}_{N} =&~ \Delta^{(\mathrm{NJ}) 2} + 4 g_{a\sigma}^{(\mathrm{NJ})} g_{\sigma a}^{(\mathrm{NJ})} N,
\end{align}
for excitation number $N \geq 1$ and propagator in the similarity frame,
\begin{align}
    \begin{aligned}
        \hat U_{\mathrm D}^{(\mathrm{NJ})}(t) =&~ e^{-i\hat{\mathcal D}^{\frac{1}{2}}\hat S_z t} \\
        =&~ \cos\left(\frac{1}{2}\hat{\mathcal D}^{\frac{1}{2}}t\right) -2i\sin\left(\frac{1}{2}\hat{\mathcal D}^{\frac{1}{2}}t\right)\hat S_z,
    \end{aligned}
\end{align}
showing trigonometric time dependence for positive real discriminant eigenvalues and hyperbolic time dependence for negative real ones; complex discriminant eigenvalues combine oscillations with exponential changes of relative amplitudes.

The second case has triangular form with $g_{a\sigma}^{(\mathrm{NJ})} g_{\sigma a}^{(\mathrm{NJ})} = 0$ and is diagonalizable, with energy eigenvalues
\begin{align}
    \begin{aligned}
        \lambda_{0} =&~ 0, \\
        \lambda_{N,\pm}^{(\mathrm{NJ})} =&~ \mp\frac{1}{2}\hbar \Delta^{(\mathrm{NJ})},
    \end{aligned}
\end{align}
and propagator in the similarity frame,
\begin{align}
    \begin{aligned}
        \hat{U}_{\mathrm{D}}^{(\mathrm{NJ})}(t) =&~ e^{-\frac{i}{\hbar}\hat{H}_{\mathrm{D}}^{(\mathrm{NJ})}t} \\
        =&~ \cos\left(\frac{1}{2}\hat{\mathcal{D}}^{\frac{1}{2}}t\right) -\frac{2i}{\hbar}\hat{H}_{\mathrm{D}}^{(\mathrm{NJ})} \hat{\mathcal{D}}_{\mathrm{P}}^{-\frac{1}{2}} \sin\left(\frac{1}{2}\hat{\mathcal{D}}^{\frac{1}{2}}t\right),
    \end{aligned}
\end{align}
showing trigonometric, hyperbolic, or mixed time dependence according to our discriminant, with no exceptional point.

The third and fourth cases have nilpotent forms with energy eigenvalues
\begin{align}
    \begin{aligned}
        \lambda_{0} =&~ 0, \\
        \lambda_{N,\pm}^{(\mathrm{NJ})} =&~ 0,
    \end{aligned}
\end{align}
and propagator in the similarity frame,
\begin{align}
    \begin{aligned}
        \hat{U}_{\mathrm{D}}^{(\mathrm{NJ})}(t) =&~ e^{-\frac{i}{\hbar}\hat{H}_{\mathrm{D}}^{(\mathrm{NJ})}t} \\
        =&~ \hat{I}-\frac{i}{\hbar}\hat{H}_{\mathrm{D}}^{(\mathrm{NJ})}t,
    \end{aligned}
\end{align}
showing linear time dependence at a Hamiltonian exceptional point when our effective Hamiltonian is nonzero.
The third case has nonzero Jordan form proportional to $\hat{S}_{+}$, with $\hat{S}_{\pm}^{2} = 0$, while in the fourth case both time-independent factors are the identity and our effective Hamiltonian $\hat{H}_{\mathrm{D}}^{(\mathrm{NJ})} = \hat{H}_{\mathrm{S}}^{(\mathrm{NJ})}$ has anti-diagonal form.
If both exchange coefficients $g_{\sigma a}^{(\mathrm{NJ})}$ and $g_{a\sigma}^{(\mathrm{NJ})}$ vanish in the fourth case, our effective Hamiltonian $\hat{H}_{\mathrm{D}}^{(\mathrm{NJ})}$ is zero and our propagator $\hat{U}_{\mathrm{D}}^{(\mathrm{NJ})}(t)$ is the identity and there is no exceptional point.

Our conditioned Hamiltonian in the original frame has energy eigenvalues
\begin{align}
    \begin{aligned}
        E_{0}^{(\mathrm{NJ})} =&~ \hbar \left[ \omega_{0}^{(\mathrm{NJ})} - \frac{1}{2} \Delta^{(\mathrm{NJ})} \right], \\
        E_{N,\pm}^{(\mathrm{NJ})} =&~ \hbar \left[ \omega_{0}^{(\mathrm{NJ})} + \omega^{(\mathrm{NJ})} N \pm \frac{1}{2} \sqrt{\mathcal{D}_{N}} \right],
    \end{aligned}
\end{align}
and conditioned propagator,
\begin{align}
    \begin{aligned}
        \hat U_{\mathrm{JC}}^{(\mathrm{NJ})}(t) =&~ e^{-\frac{i}{\hbar}\hat H_{\mathrm{JC}}^{(\mathrm{NJ})}t} \\
        =&~ e^{-\frac{i}{\hbar}\hat H_0^{(\mathrm{NJ})}t} \left[\cos\left(\frac{1}{2}\hat{\mathcal D}^{\frac{1}{2}}t\right) -\frac{i}{\hbar}\hat H_{\mathrm S}^{(\mathrm{NJ})}t\operatorname{sinc}\left(\frac{1}{2}\hat{\mathcal D}^{\frac{1}{2}}t\right)\right],
    \end{aligned}
\end{align}
valid in all four cases on the full Hilbert space of the Jaynes--Cummings model, including the zero-excitation state, with the cardinal sine function $\operatorname{sinc}(x) = \sin (x) / x$ and $\operatorname{sinc} 0 = 1$.

\section{Discussion at $T=0$}
\label{sec:Sec6}

As a practical example, we focus on zero temperature, $T = 0$, with positive coupling strength $g$, leading to vanishing thermal offset $\omega_{T} = 0$, and reservoir coefficients $\bm{\Gamma} = \bm{\Gamma}^{(+)}$ and $\bm{S} = \bm{S}^{(+)}$.
Under these conditions, the real and imaginary parts of our zero-discriminant condition at nonzero excitation number $N \geq 1$ translate into two requirements,
\begin{align}
    \begin{aligned}
        \operatorname{Re}^{2}\left[ \Delta^{(\mathrm{NJ})} \right] - \frac{1}{4} \Delta\Gamma^{2} + N\left[ 4\left\lvert g + S_{a\sigma} \right \rvert^{2} - \left\lvert \Gamma_{a\sigma} \right \rvert^{2} \right] =&~ 0, \\
        \operatorname{Re}\left[ \Delta^{(\mathrm{NJ})} \right] \Delta\Gamma + 4N\operatorname{Re}\left[ \left( g + S_{\sigma a} \right)\Gamma_{a\sigma} \right] =&~ 0,
    \end{aligned}
\end{align}
that balance shifted detuning, loss imbalance, and coherent and dissipative exchange and define an exceptional point when our normalized spin Hamiltonian $\hat{H}_{\mathrm{S}}^{(\mathrm{NJ})}$ is nonzero.

For independent reservoirs, $S_{a\sigma} = \Gamma_{a\sigma} = 0$, an exceptional point requires shifted resonance and a nonzero loss imbalance,
\begin{align}
    \begin{aligned}
        \operatorname{Re}\left[ \Delta^{(\mathrm{NJ})} \right] =&~ 0, \\
        \left\lvert \Delta\Gamma \right \rvert =&~ 4g\sqrt{N},
    \end{aligned}
\end{align}
so fixed coupling and reservoir coefficients select at most one excitation number.
Equal independent losses produce no exceptional point as $\Delta\Gamma = 0$, while loss through either component alone at rate $\Gamma_{\mu\mu}$ has threshold $\Gamma_{\mu\mu}/g = 4\sqrt{N}$.
The single-excitation condition agrees with the observed dissipative qubit--resonator threshold~\cite{Han2025p2400446}.

A common reservoir modifies both zero-discriminant conditions through the coherent-exchange coefficient $g + S_{a\sigma}$ and cross-decay coefficient $\Gamma_{a\sigma}$.
For example, eigenvalue coalescence at distinct excitation numbers with the same coupling and reservoir coefficients requires both terms of our discriminant to vanish,
\begin{align}
    \begin{aligned}
        \Delta^{(\mathrm{NJ})} =&~ 0, \\
        \operatorname{Re}\left[ \Gamma_{a\sigma} \right] =&~ \mp 2\operatorname{Im}\left[ S_{a\sigma} \right], \\
        \operatorname{Im}\left[ \Gamma_{a\sigma} \right] =&~ \pm 2\left( g + \operatorname{Re}\left[ S_{a\sigma} \right] \right),
    \end{aligned}
\end{align}
with paired signs and $g + S_{a\sigma} \neq 0$ for a nonzero nilpotent Hamiltonian.
Zero complex detuning requires equal diagonal rates $\Gamma_{aa} = \Gamma_{\sigma\sigma}$ and shifted resonance $\operatorname{Re}\left[ \Delta^{(\mathrm{NJ})} \right] = 0$, while positivity bounds the cross-decay magnitude by $\left\lvert \Gamma_{a\sigma} \right\rvert \leq \Gamma_{aa}$.
Our spectral-density matrix $\bm{J}(\nu')$ constrains both the decay matrix $\bm{\Gamma}$ and Lamb-shift matrix $\bm{S}$.
A vanishing cross Lamb shift $S_{a\sigma} = 0$ requires the principal-value integral of $J_{a\sigma}(\nu')$ to vanish.
One sufficient condition is symmetry about the carrier frequency $\omega$, $J_{a\sigma}(\omega + x) = J_{a\sigma}(\omega - x)$, with symmetric support within $\nu' > 0$, which cancels the contributions at opposite frequency offsets $x$ and $-x$.
Related experiments with two superconducting qubits in a transmission line demonstrate collective decay at one-wavelength separation, where the model describing the measured spectra has vanishing reservoir-mediated coherent exchange~\cite{vanLoo2013p1494}.

\begin{figure}[t]
    \centering
    \includegraphics[width=0.75\columnwidth]{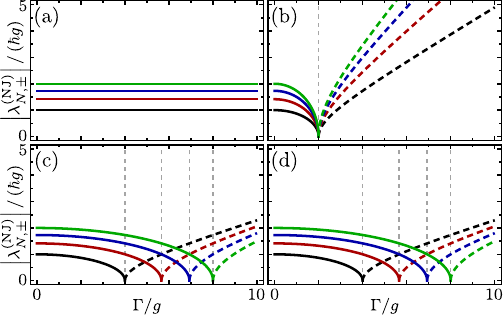}
    \caption{
Normalized magnitude of the similarity-frame energy eigenvalues, $\lvert \lambda_{N,\pm}^{(\mathrm{NJ})} \rvert/(\hbar g)$, for four loss models on shifted resonance, $\operatorname{Re}\left[ \Delta^{(\mathrm{NJ})} \right] = 0$, with zero cross Lamb shifts, $S_{a\sigma} = S_{\sigma a} = 0$.
The excitation numbers are $N = 1$ (black), $N = 2$ (red), $N = 3$ (blue), and $N = 4$ (green).
Solid and dashed curves indicate real and purely imaginary similarity-frame eigenvalues, respectively.
    (a) Equal independent losses.
    (b) Equal shared losses with relative coupling phase $\phi = \pi/2$.
    (c) Bosonic loss only.
    (d) Two-level-system loss only.
The reservoir coefficients and coherent coupling strength $g$ are identical for all excitation numbers at each loss-to-coupling ratio $\Gamma/g$.
    }
    \label{fig:Fig2}
\end{figure}

We compare four loss models on shifted resonance, $\operatorname{Re}\left[ \Delta^{(\mathrm{NJ})} \right] = 0$, with zero cross Lamb shifts, $S_{a\sigma} = S_{\sigma a} = 0$, in Fig.~\ref{fig:Fig2}.
The curves show the normalized magnitudes of our similarity-frame energy eigenvalues, $\lvert \lambda_{N,\pm}^{(\mathrm{NJ})} \rvert/(\hbar g)$, versus the loss-to-coupling ratio $\Gamma/g$ at fixed excitation number $N$.
Equal independent losses, Fig.~\ref{fig:Fig2}(a), with diagonal decay rates $\Gamma_{aa} = \Gamma_{\sigma\sigma} = \Gamma$, produce no exceptional point.
Equal shared losses, Fig.~\ref{fig:Fig2}(b), with the same diagonal decay rates and maximal cross-decay coefficient $\Gamma_{a\sigma} = i\Gamma$, corresponding to relative coupling phase $\phi = \pi/2$, produce a collective of exceptional points of order two at the common threshold $\Gamma/g = 2$.
Bosonic loss alone, Fig.~\ref{fig:Fig2}(c), and two-level-system loss alone, Fig.~\ref{fig:Fig2}(d), each produce a single exceptional point of order two at the excitation-dependent threshold $\Gamma/g = 4\sqrt{N}$.

\begin{figure}[!htbp]
    \centering
    \includegraphics[width=0.75\columnwidth]{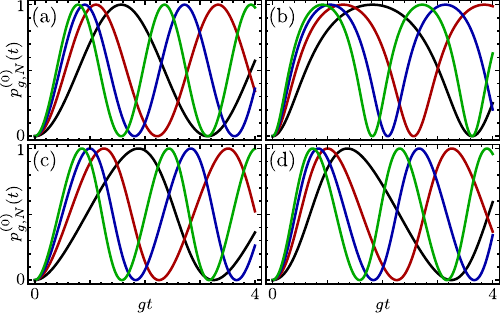}
    \caption{Conditional population of the final state $\lvert N,g\rangle$ from the initial state $\lvert N-1,e\rangle$ at loss-to-coupling ratio $\Gamma/g=1$, below the shared-loss exceptional point.
Loss models, common conditions, and excitation-number colors as in Fig.~\ref{fig:Fig2}.}
    \label{fig:Fig3}
\end{figure}

\begin{figure}[!htbp]
    \centering
    \includegraphics[width=0.75\columnwidth]{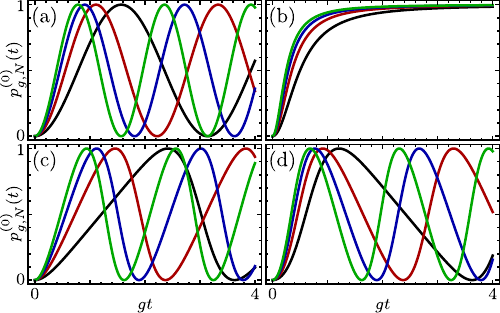}
    \caption{Conditional population of the final state $\lvert N,g\rangle$ from the initial state $\lvert N-1,e\rangle$ at loss-to-coupling ratio $\Gamma/g=2$, at the shared-loss exceptional point.
Loss models, common conditions, and excitation-number colors as in Fig.~\ref{fig:Fig2}.}
    \label{fig:Fig4}
\end{figure}

\begin{figure}[!htbp]
    \centering
    \includegraphics[width=0.75\columnwidth]{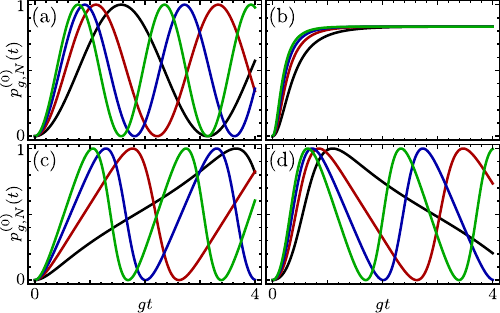}
    \caption{Conditional population of the final state $\lvert N,g\rangle$ from the initial state $\lvert N-1,e\rangle$ at loss-to-coupling ratio $\Gamma/g=3$, above the shared-loss exceptional point.
Loss models, common conditions, and excitation-number colors as in Fig.~\ref{fig:Fig2}.}
    \label{fig:Fig5}
\end{figure}

We consider the evolution of the initial state with boson number $N - 1$  and excited two-level system,
\begin{align}
    \begin{aligned}
        \lvert \psi_{N}(t) \rangle =&~ e^{-i\left( \omega_{0}^{(\mathrm{NJ})} + \omega^{(\mathrm{NJ})}N \right)t} \left[ c_{e,N}(t)\lvert N - 1,e \rangle + c_{g,N}(t)\lvert N,g \rangle \right],
    \end{aligned}
\end{align}
with dimensionless amplitudes,
\begin{align}
    \begin{aligned}
        c_{e,N}(t) =&~ \cos \left( \frac{1}{2}\sqrt{\mathcal{D}_{N}}t \right) - \frac{i}{2}\Delta^{(\mathrm{NJ})}t\operatorname{sinc}\left( \frac{1}{2}\sqrt{\mathcal{D}_{N}}t \right), \\
        c_{g,N}(t) =&~ -i g_{a\sigma}^{(\mathrm{NJ})}\sqrt{N}t \operatorname{sinc}\left( \frac{1}{2}\sqrt{\mathcal{D}_{N}}t \right),
    \end{aligned}
\end{align}
and we show the conditional population of the state with boson number $N$ and the two-level system in its ground state,
\begin{align}
    p_{g,N}^{(0)}(t) =&~ \frac{\left| c_{g,N}(t) \right|^{2}}{\left| c_{e,N}(t) \right|^{2} + \left| c_{g,N}(t) \right|^{2}},
\end{align}
versus dimensionless time $gt$ for excitation numbers $N = 1,2,3,4$ and loss-to-coupling ratios $\Gamma/g = 1$ below the shared-loss exceptional point in Fig.~\ref{fig:Fig3}, $\Gamma/g = 2$ at the exceptional point in Fig.~\ref{fig:Fig4}, and $\Gamma/g = 3$ above it in Fig.~\ref{fig:Fig5}.

Below the shared-loss exceptional point, equal independent losses, Fig.~\ref{fig:Fig3}(a), preserve the oscillatory conditional population $p_{g,N}^{(0)}(t) = \sin^{2}(g\sqrt{N}t)$.
Equal shared losses, Fig.~\ref{fig:Fig3}(b), reduce the frequency splitting and delay the first maximum while retaining complete conditional transfer.
Bosonic loss alone, Fig.~\ref{fig:Fig3}(c), favors the initial state and delays the first transfer maximum, while two-level-system loss alone, Fig.~\ref{fig:Fig3}(d), favors the final state and advances it.

At the shared-loss exceptional point, equal independent losses, Fig.~\ref{fig:Fig4}(a), retain their oscillations.
Equal shared losses, Fig.~\ref{fig:Fig4}(b), produce monotonic transfer toward unit conditional population,
\begin{align}
    p_{g,N}^{(0)}(t) =&~ \frac{4Ng^{2}t^{2}}{1 + 4Ng^{2}t^{2}},
\end{align}
with faster transfer at larger excitation number $N$.
The no-jump probability,
\begin{align}
    \begin{aligned}
        P_{N}^{(0)}(t) =&~ \langle \psi_{N}(t) \vert \psi_{N}(t) \rangle \\
        =&~ e^{-2Ngt}\left( 1 + 4Ng^{2}t^{2} \right),
    \end{aligned}
\end{align}
sets the accepted fraction~\cite{Minganti2020p062112} and has long-time limit $P_{N}^{(0)}(\infty) = 0$.
At a fixed target conditional population, this fraction decreases with excitation number despite the faster transfer.
Bosonic loss alone, Fig.~\ref{fig:Fig4}(c), and two-level-system loss alone, Fig.~\ref{fig:Fig4}(d), remain oscillatory below their excitation-dependent exceptional-point thresholds.

Above the shared-loss exceptional point, equal independent losses, Fig.~\ref{fig:Fig5}(a), retain their oscillations.
Equal shared losses, Fig.~\ref{fig:Fig5}(b), have conditional population and no-jump probability,
\begin{align}
    \begin{aligned}
        p_{g,N}^{(0)}(t) =&~ \frac{\left( \Gamma + 2g \right)\left[ \cosh\left( \sqrt{-\mathcal{D}_{N}}t \right) - 1 \right]}{2\left[ \Gamma\cosh\left( \sqrt{-\mathcal{D}_{N}}t \right) - 2g \right]}, \\
        P_{N}^{(0)}(t) =&~ e^{-\Gamma Nt}\frac{\Gamma\cosh\left( \sqrt{-\mathcal{D}_{N}}t \right) - 2g}{\Gamma - 2g},
    \end{aligned}
\end{align}
with discriminant $\mathcal{D}_{N} = N\left( 4g^{2} - \Gamma^{2} \right) < 0$.
No-jump conditioning selects the more slowly decaying eigenstate, with limiting conditional population $p_{g,N}^{(0)}(\infty) = 1/2 + g/\Gamma = 5/6$, independent of excitation number, and limiting accepted fraction $P_{N}^{(0)}(\infty) = 0$.
Bosonic loss alone, Fig.~\ref{fig:Fig5}(c), further delays the first transfer maximum, while two-level-system loss alone, Fig.~\ref{fig:Fig5}(d), further advances it, with both models still below their exceptional-point thresholds.

Our common threshold could support conditional preparation of bosonic number states through monotonic excitation transfer at one loss-to-coupling ratio for different excitation numbers.
Connecting the qubit--resonator experiment~\cite{Han2025p2400446}, collective-decay experiment~\cite{vanLoo2013p1494}, cavity--emitter proposal~\cite{Wang2019p053809}, or trapped-ion experiments on Jaynes--Cummings spin--motion coupling~\cite{Meekhof1996p1796} and tunable motional damping~\cite{So2024peads8011} to our required decay matrix $\bm{\Gamma}$ and Lamb-shift matrix $\bm{S}$ remains conjectural.
We also conjecture that Jaynes--Cummings systems with chiral coupling~\cite{Junge2013p213604,Scheucher2016p1577} may provide an alternative route to realizing our required decay and Lamb-shift coefficients.

\section{Conclusion}
\label{sec:Sec7}

We introduced a Moore--Penrose normalized $\mathrm{su}(2)$ representation that diagonalizes the Jaynes--Cummings Hamiltonian through a single operator rotation on the full Hilbert space and provides its global spectral resolution.

Using our spectral resolution, we derived microscopic master equations for the Jaynes--Cummings system coupled to a common bosonic reservoir at finite temperature, $T \geq 0$, under full and partial Bohr-frequency resolution.
For bosonic loss alone, neglecting Lamb shifts, we recover the fully resolved microscopic cavity-loss master equations based on sector-by-sector diagonalization~\cite{Scala2007p013811,Gonzalez2018p015301}.
Our treatment generalizes these descriptions and the finite-temperature global operator treatment of cavity loss~\cite{OzakiNakazato2021} and includes bosonic loss alone, two-level-system loss alone, and losses through both components, with correlations ranging continuously from independent to maximally correlated reservoir channels.
Our fully resolved conditioned Hamiltonian is diagonal in the global dressed basis and has no exceptional points.

Under partial Bohr-frequency resolution, we showed that our conditioned Hamiltonian retains interference between unresolved transitions and can exhibit exceptional points.
We derived its spectrum and exceptional-point conditions and provided a unified conditioned propagator for diagonalizable and non-diagonalizable cases on the full Hilbert space.

At zero temperature on shifted resonance, we recovered the phenomenological exceptional-point threshold for bosonic loss alone, including the observed single-excitation value~\cite{Han2025p2400446}, and predicted the same excitation-dependent threshold for two-level-system loss alone.
We showed that equal independent losses produce no exceptional point.
For equal shared losses with vanishing cross Lamb shifts and a relative coupling phase of $\pi/2$, we predicted a collective of exceptional points of order two with a common threshold independent of excitation number.

Our microscopic theory connects reservoir correlations to conditioned spectra and dynamics, providing a basis for assessing when phenomenological exceptional-point models apply.
Our common threshold could support conditional preparation of bosonic number states at one loss-to-coupling ratio for different excitation numbers.
Realizing our required decay and Lamb-shift coefficients in superconducting systems~\cite{Han2025p2400446,vanLoo2013p1494}, cavity--emitter systems~\cite{Wang2019p053809}, or trapped-ion systems~\cite{Meekhof1996p1796,So2024peads8011} remains conjectural.
Jaynes--Cummings systems with chiral coupling~\cite{Junge2013p213604,Scheucher2016p1577} may provide an alternative route.

\section*{Funding}
The author received no external funding for this work.

\begin{acknowledgments}
B.~M.~R.~L. thanks Jacinta Alderete Galan for providing daycare support throughout this work.
He also acknowledges support and hospitality as an affiliate visiting colleague at the Department of Physics and Astronomy, University of New Mexico, and thanks Andreas Hanke at the University of Texas Rio Grande Valley for useful discussion.
He also thanks Anuar Kafuri Zarzosa, F\'{e}lix Humberto Maldonado Villamizar, and Julio Abraham Mendoza Fierro  for fruitful discussions that improved the manuscript during review.
\end{acknowledgments}

\section*{Disclosures}
The author declares no conflicts of interest.

\section*{Data Availability Statement}
No data were generated or analyzed in this theoretical study.



\begin{thebibliography}{45}%
	\makeatletter
	\providecommand \@ifxundefined [1]{%
		\@ifx{#1\undefined}
	}%
	\providecommand \@ifnum [1]{%
		\ifnum #1\expandafter \@firstoftwo
		\else \expandafter \@secondoftwo
		\fi
	}%
	\providecommand \@ifx [1]{%
		\ifx #1\expandafter \@firstoftwo
		\else \expandafter \@secondoftwo
		\fi
	}%
	\providecommand \natexlab [1]{#1}%
	\providecommand \enquote  [1]{``#1''}%
	\providecommand \bibnamefont  [1]{#1}%
	\providecommand \bibfnamefont [1]{#1}%
	\providecommand \citenamefont [1]{#1}%
	\providecommand \href@noop [0]{\@secondoftwo}%
	\providecommand \href [0]{\begingroup \@sanitize@url \@href}%
	\providecommand \@href[1]{\@@startlink{#1}\@@href}%
	\providecommand \@@href[1]{\endgroup#1\@@endlink}%
	\providecommand \@sanitize@url [0]{\catcode `\\12\catcode `\$12\catcode
		`\&12\catcode `\#12\catcode `\^12\catcode `\_12\catcode `\%12\relax}%
	\providecommand \@@startlink[1]{}%
	\providecommand \@@endlink[0]{}%
	\providecommand \url  [0]{\begingroup\@sanitize@url \@url }%
	\providecommand \@url [1]{\endgroup\@href {#1}{\urlprefix }}%
	\providecommand \urlprefix  [0]{URL }%
	\providecommand \Eprint [0]{\href }%
	\providecommand \doibase [0]{https://doi.org/}%
	\providecommand \selectlanguage [0]{\@gobble}%
	\providecommand \bibinfo  [0]{\@secondoftwo}%
	\providecommand \bibfield  [0]{\@secondoftwo}%
	\providecommand \translation [1]{[#1]}%
	\providecommand \BibitemOpen [0]{}%
	\providecommand \bibitemStop [0]{}%
	\providecommand \bibitemNoStop [0]{.\EOS\space}%
	\providecommand \EOS [0]{\spacefactor3000\relax}%
	\providecommand \BibitemShut  [1]{\csname bibitem#1\endcsname}%
	\let\auto@bib@innerbib\@empty
	\bibitem [{\citenamefont {Kato}(1966)}]{Kato1966}%
	\BibitemOpen
	\bibfield  {author} {\bibinfo {author} {\bibfnamefont {T.}~\bibnamefont
			{Kato}},\ }\href {https://doi.org/10.1007/978-3-662-12678-3} {\emph {\bibinfo
			{title} {Perturbation theory for linear operators}}},\ \bibinfo {series}
	{Grundlehren der mathematischen Wissenschaften}, Vol.\ \bibinfo {volume}
	{132}\ (\bibinfo  {publisher} {Springer},\ \bibinfo {address} {Berlin},\
	\bibinfo {year} {1966})\BibitemShut {NoStop}%
	\bibitem [{\citenamefont {Heiss}\ and\ \citenamefont
		{Sannino}(1990)}]{Heiss1990p1167}%
	\BibitemOpen
	\bibfield  {author} {\bibinfo {author} {\bibfnamefont {W.~D.}\ \bibnamefont
			{Heiss}}\ and\ \bibinfo {author} {\bibfnamefont {A.~L.}\ \bibnamefont
			{Sannino}},\ }\bibfield  {title} {\bibinfo {title} {Avoided level crossing
			and exceptional points},\ }\href {https://doi.org/10.1088/0305-4470/23/7/022}
	{\bibfield  {journal} {\bibinfo  {journal} {J. Phys. A: Math. Gen.}\ }\textbf
		{\bibinfo {volume} {23}},\ \bibinfo {pages} {1167} (\bibinfo {year}
		{1990})}\BibitemShut {NoStop}%
	\bibitem [{\citenamefont {Heiss}(2012)}]{Heiss2012p444016}%
	\BibitemOpen
	\bibfield  {author} {\bibinfo {author} {\bibfnamefont {W.~D.}\ \bibnamefont
			{Heiss}},\ }\bibfield  {title} {\bibinfo {title} {The physics of exceptional
			points},\ }\href {https://doi.org/10.1088/1751-8113/45/44/444016} {\bibfield
		{journal} {\bibinfo  {journal} {J. Phys. A: Math. Theor.}\ }\textbf {\bibinfo
			{volume} {45}},\ \bibinfo {pages} {444016} (\bibinfo {year}
		{2012})}\BibitemShut {NoStop}%
	\bibitem [{\citenamefont {Dembowski}\ \emph {et~al.}(2001)\citenamefont
		{Dembowski}, \citenamefont {Gr{\"a}f}, \citenamefont {Harney}, \citenamefont
		{Heine}, \citenamefont {Heiss},\ and\ \citenamefont
		{Richter}}]{Dembowski2001p787}%
	\BibitemOpen
	\bibfield  {author} {\bibinfo {author} {\bibfnamefont {C.}~\bibnamefont
			{Dembowski}}, \bibinfo {author} {\bibfnamefont {H.-D.}\ \bibnamefont
			{Gr{\"a}f}}, \bibinfo {author} {\bibfnamefont {H.~L.}\ \bibnamefont
			{Harney}}, \bibinfo {author} {\bibfnamefont {A.}~\bibnamefont {Heine}},
		\bibinfo {author} {\bibfnamefont {W.~D.}\ \bibnamefont {Heiss}},\ and\
		\bibinfo {author} {\bibfnamefont {A.}~\bibnamefont {Richter}},\ }\bibfield
	{title} {\bibinfo {title} {Experimental observation of the topological
			structure of exceptional points},\ }\href
	{https://doi.org/10.1103/PhysRevLett.86.787} {\bibfield  {journal} {\bibinfo
			{journal} {Phys. Rev. Lett.}\ }\textbf {\bibinfo {volume} {86}},\ \bibinfo
		{pages} {787} (\bibinfo {year} {2001})}\BibitemShut {NoStop}%
	\bibitem [{\citenamefont {Guo}\ \emph {et~al.}(2009)\citenamefont {Guo},
		\citenamefont {Salamo}, \citenamefont {Duchesne}, \citenamefont {Morandotti},
		\citenamefont {Volatier-Ravat}, \citenamefont {Aimez}, \citenamefont
		{Siviloglou},\ and\ \citenamefont {Christodoulides}}]{Guo2009p093902}%
	\BibitemOpen
	\bibfield  {author} {\bibinfo {author} {\bibfnamefont {A.}~\bibnamefont
			{Guo}}, \bibinfo {author} {\bibfnamefont {G.~J.}\ \bibnamefont {Salamo}},
		\bibinfo {author} {\bibfnamefont {D.}~\bibnamefont {Duchesne}}, \bibinfo
		{author} {\bibfnamefont {R.}~\bibnamefont {Morandotti}}, \bibinfo {author}
		{\bibfnamefont {M.}~\bibnamefont {Volatier-Ravat}}, \bibinfo {author}
		{\bibfnamefont {V.}~\bibnamefont {Aimez}}, \bibinfo {author} {\bibfnamefont
			{G.~A.}\ \bibnamefont {Siviloglou}},\ and\ \bibinfo {author} {\bibfnamefont
			{D.~N.}\ \bibnamefont {Christodoulides}},\ }\bibfield  {title} {\bibinfo
		{title} {Observation of {$\mathcal{PT}$}-symmetry breaking in complex optical
			potentials},\ }\href {https://doi.org/10.1103/PhysRevLett.103.093902}
	{\bibfield  {journal} {\bibinfo  {journal} {Phys. Rev. Lett.}\ }\textbf
		{\bibinfo {volume} {103}},\ \bibinfo {pages} {093902} (\bibinfo {year}
		{2009})}\BibitemShut {NoStop}%
	\bibitem [{\citenamefont {R{\"u}ter}\ \emph {et~al.}(2010)\citenamefont
		{R{\"u}ter}, \citenamefont {Makris}, \citenamefont {El-Ganainy},
		\citenamefont {Christodoulides}, \citenamefont {Segev},\ and\ \citenamefont
		{Kip}}]{Ruter2010p192}%
	\BibitemOpen
	\bibfield  {author} {\bibinfo {author} {\bibfnamefont {C.~E.}\ \bibnamefont
			{R{\"u}ter}}, \bibinfo {author} {\bibfnamefont {K.~G.}\ \bibnamefont
			{Makris}}, \bibinfo {author} {\bibfnamefont {R.}~\bibnamefont {El-Ganainy}},
		\bibinfo {author} {\bibfnamefont {D.~N.}\ \bibnamefont {Christodoulides}},
		\bibinfo {author} {\bibfnamefont {M.}~\bibnamefont {Segev}},\ and\ \bibinfo
		{author} {\bibfnamefont {D.}~\bibnamefont {Kip}},\ }\bibfield  {title}
	{\bibinfo {title} {Observation of parity--time symmetry in optics},\ }\href
	{https://doi.org/10.1038/nphys1515} {\bibfield  {journal} {\bibinfo
			{journal} {Nat. Phys.}\ }\textbf {\bibinfo {volume} {6}},\ \bibinfo {pages}
		{192} (\bibinfo {year} {2010})}\BibitemShut {NoStop}%
	\bibitem [{\citenamefont {Peng}\ \emph {et~al.}(2014)\citenamefont {Peng},
		\citenamefont {{\"O}zdemir}, \citenamefont {Lei}, \citenamefont {Monifi},
		\citenamefont {Gianfreda}, \citenamefont {Long}, \citenamefont {Fan},
		\citenamefont {Nori}, \citenamefont {Bender},\ and\ \citenamefont
		{Yang}}]{Peng2014p394}%
	\BibitemOpen
	\bibfield  {author} {\bibinfo {author} {\bibfnamefont {B.}~\bibnamefont
			{Peng}}, \bibinfo {author} {\bibfnamefont {{\c{S}}.~K.}\ \bibnamefont
			{{\"O}zdemir}}, \bibinfo {author} {\bibfnamefont {F.}~\bibnamefont {Lei}},
		\bibinfo {author} {\bibfnamefont {F.}~\bibnamefont {Monifi}}, \bibinfo
		{author} {\bibfnamefont {M.}~\bibnamefont {Gianfreda}}, \bibinfo {author}
		{\bibfnamefont {G.~L.}\ \bibnamefont {Long}}, \bibinfo {author}
		{\bibfnamefont {S.}~\bibnamefont {Fan}}, \bibinfo {author} {\bibfnamefont
			{F.}~\bibnamefont {Nori}}, \bibinfo {author} {\bibfnamefont {C.~M.}\
			\bibnamefont {Bender}},\ and\ \bibinfo {author} {\bibfnamefont
			{L.}~\bibnamefont {Yang}},\ }\bibfield  {title} {\bibinfo {title}
		{Parity--time-symmetric whispering-gallery microcavities},\ }\href
	{https://doi.org/10.1038/nphys2927} {\bibfield  {journal} {\bibinfo
			{journal} {Nat. Phys.}\ }\textbf {\bibinfo {volume} {10}},\ \bibinfo {pages}
		{394} (\bibinfo {year} {2014})}\BibitemShut {NoStop}%
	\bibitem [{\citenamefont {Quiroz-Ju{\'a}rez}\ \emph {et~al.}(2019)\citenamefont
		{Quiroz-Ju{\'a}rez}, \citenamefont {Perez-Leija}, \citenamefont {Tschernig},
		\citenamefont {Rodr{\'i}guez-Lara}, \citenamefont {Maga{\~n}a-Loaiza},
		\citenamefont {Busch}, \citenamefont {Joglekar},\ and\ \citenamefont
		{de~J.~Le{\'o}n-Montiel}}]{QuirozJuarez2019p862}%
	\BibitemOpen
	\bibfield  {author} {\bibinfo {author} {\bibfnamefont {M.~A.}\ \bibnamefont
			{Quiroz-Ju{\'a}rez}}, \bibinfo {author} {\bibfnamefont {A.}~\bibnamefont
			{Perez-Leija}}, \bibinfo {author} {\bibfnamefont {K.}~\bibnamefont
			{Tschernig}}, \bibinfo {author} {\bibfnamefont {B.~M.}\ \bibnamefont
			{Rodr{\'i}guez-Lara}}, \bibinfo {author} {\bibfnamefont {O.~S.}\ \bibnamefont
			{Maga{\~n}a-Loaiza}}, \bibinfo {author} {\bibfnamefont {K.}~\bibnamefont
			{Busch}}, \bibinfo {author} {\bibfnamefont {Y.~N.}\ \bibnamefont
			{Joglekar}},\ and\ \bibinfo {author} {\bibfnamefont {R.}~\bibnamefont
			{de~J.~Le{\'o}n-Montiel}},\ }\bibfield  {title} {\bibinfo {title}
		{Exceptional points of any order in a single, lossy waveguide beam splitter
			by photon-number-resolved detection},\ }\href
	{https://doi.org/10.1364/PRJ.7.000862} {\bibfield  {journal} {\bibinfo
			{journal} {Photon. Res.}\ }\textbf {\bibinfo {volume} {7}},\ \bibinfo {pages}
		{862} (\bibinfo {year} {2019})}\BibitemShut {NoStop}%
	\bibitem [{\citenamefont {Naghiloo}\ \emph {et~al.}(2019)\citenamefont
		{Naghiloo}, \citenamefont {Abbasi}, \citenamefont {Joglekar},\ and\
		\citenamefont {Murch}}]{Naghiloo2019p1232}%
	\BibitemOpen
	\bibfield  {author} {\bibinfo {author} {\bibfnamefont {M.}~\bibnamefont
			{Naghiloo}}, \bibinfo {author} {\bibfnamefont {M.}~\bibnamefont {Abbasi}},
		\bibinfo {author} {\bibfnamefont {Y.~N.}\ \bibnamefont {Joglekar}},\ and\
		\bibinfo {author} {\bibfnamefont {K.~W.}\ \bibnamefont {Murch}},\ }\bibfield
	{title} {\bibinfo {title} {Quantum state tomography across the exceptional
			point in a single dissipative qubit},\ }\href
	{https://doi.org/10.1038/s41567-019-0652-z} {\bibfield  {journal} {\bibinfo
			{journal} {Nat. Phys.}\ }\textbf {\bibinfo {volume} {15}},\ \bibinfo {pages}
		{1232} (\bibinfo {year} {2019})}\BibitemShut {NoStop}%
	\bibitem [{\citenamefont {Wiersig}(2014)}]{Wiersig2014p203901}%
	\BibitemOpen
	\bibfield  {author} {\bibinfo {author} {\bibfnamefont {J.}~\bibnamefont
			{Wiersig}},\ }\bibfield  {title} {\bibinfo {title} {Enhancing the sensitivity
			of frequency and energy splitting detection by using exceptional points:
			Application to microcavity sensors for single-particle detection},\ }\href
	{https://doi.org/10.1103/PhysRevLett.112.203901} {\bibfield  {journal}
		{\bibinfo  {journal} {Phys. Rev. Lett.}\ }\textbf {\bibinfo {volume} {112}},\
		\bibinfo {pages} {203901} (\bibinfo {year} {2014})}\BibitemShut {NoStop}%
	\bibitem [{\citenamefont {Chen}\ \emph {et~al.}(2017)\citenamefont {Chen},
		\citenamefont {{\"O}zdemir}, \citenamefont {Zhao}, \citenamefont {Wiersig},\
		and\ \citenamefont {Yang}}]{Chen2017p192}%
	\BibitemOpen
	\bibfield  {author} {\bibinfo {author} {\bibfnamefont {W.}~\bibnamefont
			{Chen}}, \bibinfo {author} {\bibfnamefont {{\c{S}}.~K.}\ \bibnamefont
			{{\"O}zdemir}}, \bibinfo {author} {\bibfnamefont {G.}~\bibnamefont {Zhao}},
		\bibinfo {author} {\bibfnamefont {J.}~\bibnamefont {Wiersig}},\ and\ \bibinfo
		{author} {\bibfnamefont {L.}~\bibnamefont {Yang}},\ }\bibfield  {title}
	{\bibinfo {title} {Exceptional points enhance sensing in an optical
			microcavity},\ }\href {https://doi.org/10.1038/nature23281} {\bibfield
		{journal} {\bibinfo  {journal} {Nature}\ }\textbf {\bibinfo {volume} {548}},\
		\bibinfo {pages} {192} (\bibinfo {year} {2017})}\BibitemShut {NoStop}%
	\bibitem [{\citenamefont {Hodaei}\ \emph {et~al.}(2017)\citenamefont {Hodaei},
		\citenamefont {Hassan}, \citenamefont {Wittek}, \citenamefont
		{{Garcia-Gracia}}, \citenamefont {El-Ganainy}, \citenamefont
		{Christodoulides},\ and\ \citenamefont {Khajavikhan}}]{Hodaei2017p187}%
	\BibitemOpen
	\bibfield  {author} {\bibinfo {author} {\bibfnamefont {H.}~\bibnamefont
			{Hodaei}}, \bibinfo {author} {\bibfnamefont {A.~U.}\ \bibnamefont {Hassan}},
		\bibinfo {author} {\bibfnamefont {S.}~\bibnamefont {Wittek}}, \bibinfo
		{author} {\bibfnamefont {H.}~\bibnamefont {{Garcia-Gracia}}}, \bibinfo
		{author} {\bibfnamefont {R.}~\bibnamefont {El-Ganainy}}, \bibinfo {author}
		{\bibfnamefont {D.~N.}\ \bibnamefont {Christodoulides}},\ and\ \bibinfo
		{author} {\bibfnamefont {M.}~\bibnamefont {Khajavikhan}},\ }\bibfield
	{title} {\bibinfo {title} {Enhanced sensitivity at higher-order exceptional
			points},\ }\href {https://doi.org/10.1038/nature23280} {\bibfield  {journal}
		{\bibinfo  {journal} {Nature}\ }\textbf {\bibinfo {volume} {548}},\ \bibinfo
		{pages} {187} (\bibinfo {year} {2017})}\BibitemShut {NoStop}%
	\bibitem [{\citenamefont {Jaynes}\ and\ \citenamefont
		{Cummings}(1963)}]{Jaynes1963p89}%
	\BibitemOpen
	\bibfield  {author} {\bibinfo {author} {\bibfnamefont {E.~T.}\ \bibnamefont
			{Jaynes}}\ and\ \bibinfo {author} {\bibfnamefont {F.~W.}\ \bibnamefont
			{Cummings}},\ }\bibfield  {title} {\bibinfo {title} {Comparison of quantum
			and semiclassical radiation theories with application to the beam maser},\
	}\href {https://doi.org/10.1109/PROC.1963.1664} {\bibfield  {journal}
		{\bibinfo  {journal} {Proc. IEEE}\ }\textbf {\bibinfo {volume} {51}},\
		\bibinfo {pages} {89} (\bibinfo {year} {1963})}\BibitemShut {NoStop}%
	\bibitem [{\citenamefont {Thompson}\ \emph {et~al.}(1992)\citenamefont
		{Thompson}, \citenamefont {Rempe},\ and\ \citenamefont
		{Kimble}}]{Thompson1992p1132}%
	\BibitemOpen
	\bibfield  {author} {\bibinfo {author} {\bibfnamefont {R.~J.}\ \bibnamefont
			{Thompson}}, \bibinfo {author} {\bibfnamefont {G.}~\bibnamefont {Rempe}},\
		and\ \bibinfo {author} {\bibfnamefont {H.~J.}\ \bibnamefont {Kimble}},\
	}\bibfield  {title} {\bibinfo {title} {Observation of normal-mode splitting
			for an atom in an optical cavity},\ }\href
	{https://doi.org/10.1103/PhysRevLett.68.1132} {\bibfield  {journal} {\bibinfo
			{journal} {Phys. Rev. Lett.}\ }\textbf {\bibinfo {volume} {68}},\ \bibinfo
		{pages} {1132} (\bibinfo {year} {1992})}\BibitemShut {NoStop}%
	\bibitem [{\citenamefont {Leibfried}\ \emph {et~al.}(2003)\citenamefont
		{Leibfried}, \citenamefont {Blatt}, \citenamefont {Monroe},\ and\
		\citenamefont {Wineland}}]{Leibfried2003p281}%
	\BibitemOpen
	\bibfield  {author} {\bibinfo {author} {\bibfnamefont {D.}~\bibnamefont
			{Leibfried}}, \bibinfo {author} {\bibfnamefont {R.}~\bibnamefont {Blatt}},
		\bibinfo {author} {\bibfnamefont {C.}~\bibnamefont {Monroe}},\ and\ \bibinfo
		{author} {\bibfnamefont {D.}~\bibnamefont {Wineland}},\ }\bibfield  {title}
	{\bibinfo {title} {Quantum dynamics of single trapped ions},\ }\href
	{https://doi.org/10.1103/RevModPhys.75.281} {\bibfield  {journal} {\bibinfo
			{journal} {Rev. Mod. Phys.}\ }\textbf {\bibinfo {volume} {75}},\ \bibinfo
		{pages} {281} (\bibinfo {year} {2003})}\BibitemShut {NoStop}%
	\bibitem [{\citenamefont {Wallraff}\ \emph {et~al.}(2004)\citenamefont
		{Wallraff}, \citenamefont {Schuster}, \citenamefont {Blais}, \citenamefont
		{Frunzio}, \citenamefont {Huang}, \citenamefont {Majer}, \citenamefont
		{Kumar}, \citenamefont {Girvin},\ and\ \citenamefont
		{Schoelkopf}}]{Wallraff2004p162}%
	\BibitemOpen
	\bibfield  {author} {\bibinfo {author} {\bibfnamefont {A.}~\bibnamefont
			{Wallraff}}, \bibinfo {author} {\bibfnamefont {D.~I.}\ \bibnamefont
			{Schuster}}, \bibinfo {author} {\bibfnamefont {A.}~\bibnamefont {Blais}},
		\bibinfo {author} {\bibfnamefont {L.}~\bibnamefont {Frunzio}}, \bibinfo
		{author} {\bibfnamefont {R.-S.}\ \bibnamefont {Huang}}, \bibinfo {author}
		{\bibfnamefont {J.}~\bibnamefont {Majer}}, \bibinfo {author} {\bibfnamefont
			{S.}~\bibnamefont {Kumar}}, \bibinfo {author} {\bibfnamefont {S.~M.}\
			\bibnamefont {Girvin}},\ and\ \bibinfo {author} {\bibfnamefont {R.~J.}\
			\bibnamefont {Schoelkopf}},\ }\bibfield  {title} {\bibinfo {title} {Strong
			coupling of a single photon to a superconducting qubit using circuit quantum
			electrodynamics},\ }\href {https://doi.org/10.1038/nature02851} {\bibfield
		{journal} {\bibinfo  {journal} {Nature}\ }\textbf {\bibinfo {volume} {431}},\
		\bibinfo {pages} {162} (\bibinfo {year} {2004})}\BibitemShut {NoStop}%
	\bibitem [{\citenamefont {Blais}\ \emph {et~al.}(2021)\citenamefont {Blais},
		\citenamefont {Grimsmo}, \citenamefont {Girvin},\ and\ \citenamefont
		{Wallraff}}]{Blais2021p025005}%
	\BibitemOpen
	\bibfield  {author} {\bibinfo {author} {\bibfnamefont {A.}~\bibnamefont
			{Blais}}, \bibinfo {author} {\bibfnamefont {A.~L.}\ \bibnamefont {Grimsmo}},
		\bibinfo {author} {\bibfnamefont {S.~M.}\ \bibnamefont {Girvin}},\ and\
		\bibinfo {author} {\bibfnamefont {A.}~\bibnamefont {Wallraff}},\ }\bibfield
	{title} {\bibinfo {title} {Circuit quantum electrodynamics},\ }\href
	{https://doi.org/10.1103/RevModPhys.93.025005} {\bibfield  {journal}
		{\bibinfo  {journal} {Rev. Mod. Phys.}\ }\textbf {\bibinfo {volume} {93}},\
		\bibinfo {pages} {025005} (\bibinfo {year} {2021})}\BibitemShut {NoStop}%
	\bibitem [{\citenamefont {Scala}\ \emph {et~al.}(2007)\citenamefont {Scala},
		\citenamefont {Militello}, \citenamefont {Messina}, \citenamefont {Piilo},\
		and\ \citenamefont {Maniscalco}}]{Scala2007p013811}%
	\BibitemOpen
	\bibfield  {author} {\bibinfo {author} {\bibfnamefont {M.}~\bibnamefont
			{Scala}}, \bibinfo {author} {\bibfnamefont {B.}~\bibnamefont {Militello}},
		\bibinfo {author} {\bibfnamefont {A.}~\bibnamefont {Messina}}, \bibinfo
		{author} {\bibfnamefont {J.}~\bibnamefont {Piilo}},\ and\ \bibinfo {author}
		{\bibfnamefont {S.}~\bibnamefont {Maniscalco}},\ }\bibfield  {title}
	{\bibinfo {title} {Microscopic derivation of the {Jaynes--Cummings} model
			with cavity losses},\ }\href {https://doi.org/10.1103/PhysRevA.75.013811}
	{\bibfield  {journal} {\bibinfo  {journal} {Phys. Rev. A}\ }\textbf {\bibinfo
			{volume} {75}},\ \bibinfo {pages} {013811} (\bibinfo {year}
		{2007})}\BibitemShut {NoStop}%
	\bibitem [{\citenamefont {Gonz\'alez-Guti\'errez}\ \emph
		{et~al.}(2018)\citenamefont {Gonz\'alez-Guti\'errez}, \citenamefont
		{Sol\'is-Valles},\ and\ \citenamefont
		{Rodr\'iguez-Lara}}]{Gonzalez2018p015301}%
	\BibitemOpen
	\bibfield  {author} {\bibinfo {author} {\bibfnamefont {C.~A.}\ \bibnamefont
			{Gonz\'alez-Guti\'errez}}, \bibinfo {author} {\bibfnamefont {D.}~\bibnamefont
			{Sol\'is-Valles}},\ and\ \bibinfo {author} {\bibfnamefont {B.~M.}\
			\bibnamefont {Rodr\'iguez-Lara}},\ }\bibfield  {title} {\bibinfo {title}
		{Microscopic approach to field dissipation in the {Jaynes--Cummings} model},\
	}\href {https://doi.org/10.1088/1751-8121/aa93c3} {\bibfield  {journal}
		{\bibinfo  {journal} {J. Phys. A: Math. Theor.}\ }\textbf {\bibinfo {volume}
			{51}},\ \bibinfo {pages} {015301} (\bibinfo {year} {2018})}\BibitemShut
	{NoStop}%
	\bibitem [{\citenamefont {Ozaki}\ and\ \citenamefont
		{Nakazato}(2021)}]{OzakiNakazato2021}%
	\BibitemOpen
	\bibfield  {author} {\bibinfo {author} {\bibfnamefont {S.}~\bibnamefont
			{Ozaki}}\ and\ \bibinfo {author} {\bibfnamefont {H.}~\bibnamefont
			{Nakazato}},\ }\bibfield  {title} {\bibinfo {title} {Analytic approach to
			dynamics of the resonant and off-resonant {Jaynes--Cummings} systems with
			cavity losses},\ }\href {https://doi.org/10.1103/PhysRevA.103.053713}
	{\bibfield  {journal} {\bibinfo  {journal} {Phys. Rev. A}\ }\textbf {\bibinfo
			{volume} {103}},\ \bibinfo {pages} {053713} (\bibinfo {year}
		{2021})}\BibitemShut {NoStop}%
	\bibitem [{\citenamefont {Dalibard}\ \emph {et~al.}(1992)\citenamefont
		{Dalibard}, \citenamefont {Castin},\ and\ \citenamefont
		{M{\o}lmer}}]{Dalibard1992p580}%
	\BibitemOpen
	\bibfield  {author} {\bibinfo {author} {\bibfnamefont {J.}~\bibnamefont
			{Dalibard}}, \bibinfo {author} {\bibfnamefont {Y.}~\bibnamefont {Castin}},\
		and\ \bibinfo {author} {\bibfnamefont {K.}~\bibnamefont {M{\o}lmer}},\
	}\bibfield  {title} {\bibinfo {title} {Wave-function approach to dissipative
			processes in quantum optics},\ }\href
	{https://doi.org/10.1103/PhysRevLett.68.580} {\bibfield  {journal} {\bibinfo
			{journal} {Phys. Rev. Lett.}\ }\textbf {\bibinfo {volume} {68}},\ \bibinfo
		{pages} {580} (\bibinfo {year} {1992})}\BibitemShut {NoStop}%
	\bibitem [{\citenamefont {Carmichael}(1993)}]{Carmichael1993p2273}%
	\BibitemOpen
	\bibfield  {author} {\bibinfo {author} {\bibfnamefont {H.~J.}\ \bibnamefont
			{Carmichael}},\ }\bibfield  {title} {\bibinfo {title} {Quantum trajectory
			theory for cascaded open systems},\ }\href
	{https://doi.org/10.1103/PhysRevLett.70.2273} {\bibfield  {journal} {\bibinfo
			{journal} {Phys. Rev. Lett.}\ }\textbf {\bibinfo {volume} {70}},\ \bibinfo
		{pages} {2273} (\bibinfo {year} {1993})}\BibitemShut {NoStop}%
	\bibitem [{\citenamefont {Plenio}\ and\ \citenamefont
		{Knight}(1998)}]{Plenio1998p101}%
	\BibitemOpen
	\bibfield  {author} {\bibinfo {author} {\bibfnamefont {M.~B.}\ \bibnamefont
			{Plenio}}\ and\ \bibinfo {author} {\bibfnamefont {P.~L.}\ \bibnamefont
			{Knight}},\ }\bibfield  {title} {\bibinfo {title} {The quantum-jump approach
			to dissipative dynamics in quantum optics},\ }\href
	{https://doi.org/10.1103/RevModPhys.70.101} {\bibfield  {journal} {\bibinfo
			{journal} {Rev. Mod. Phys.}\ }\textbf {\bibinfo {volume} {70}},\ \bibinfo
		{pages} {101} (\bibinfo {year} {1998})}\BibitemShut {NoStop}%
	\bibitem [{\citenamefont {Minganti}\ \emph {et~al.}(2019)\citenamefont
		{Minganti}, \citenamefont {Miranowicz}, \citenamefont {Chhajlany},\ and\
		\citenamefont {Nori}}]{Minganti2019p062131}%
	\BibitemOpen
	\bibfield  {author} {\bibinfo {author} {\bibfnamefont {F.}~\bibnamefont
			{Minganti}}, \bibinfo {author} {\bibfnamefont {A.}~\bibnamefont
			{Miranowicz}}, \bibinfo {author} {\bibfnamefont {R.~W.}\ \bibnamefont
			{Chhajlany}},\ and\ \bibinfo {author} {\bibfnamefont {F.}~\bibnamefont
			{Nori}},\ }\bibfield  {title} {\bibinfo {title} {Quantum exceptional points
			of non-{Hermitian} {Hamiltonians} and {Liouvillians}: the effects of quantum
			jumps},\ }\href {https://doi.org/10.1103/PhysRevA.100.062131} {\bibfield
		{journal} {\bibinfo  {journal} {Phys. Rev. A}\ }\textbf {\bibinfo {volume}
			{100}},\ \bibinfo {pages} {062131} (\bibinfo {year} {2019})}\BibitemShut
	{NoStop}%
	\bibitem [{\citenamefont {Minganti}\ \emph {et~al.}(2020)\citenamefont
		{Minganti}, \citenamefont {Miranowicz}, \citenamefont {Chhajlany},
		\citenamefont {Arkhipov},\ and\ \citenamefont {Nori}}]{Minganti2020p062112}%
	\BibitemOpen
	\bibfield  {author} {\bibinfo {author} {\bibfnamefont {F.}~\bibnamefont
			{Minganti}}, \bibinfo {author} {\bibfnamefont {A.}~\bibnamefont
			{Miranowicz}}, \bibinfo {author} {\bibfnamefont {R.~W.}\ \bibnamefont
			{Chhajlany}}, \bibinfo {author} {\bibfnamefont {I.~I.}\ \bibnamefont
			{Arkhipov}},\ and\ \bibinfo {author} {\bibfnamefont {F.}~\bibnamefont
			{Nori}},\ }\bibfield  {title} {\bibinfo {title} {Hybrid-{Liouvillian}
			formalism connecting exceptional points of non-{Hermitian} {Hamiltonians} and
			{Liouvillians} via postselection of quantum trajectories},\ }\href
	{https://doi.org/10.1103/PhysRevA.101.062112} {\bibfield  {journal} {\bibinfo
			{journal} {Phys. Rev. A}\ }\textbf {\bibinfo {volume} {101}},\ \bibinfo
		{pages} {062112} (\bibinfo {year} {2020})}\BibitemShut {NoStop}%
	\bibitem [{\citenamefont {Chen}\ \emph {et~al.}(2021)\citenamefont {Chen},
		\citenamefont {Abbasi}, \citenamefont {Joglekar},\ and\ \citenamefont
		{Murch}}]{Chen2021p140504}%
	\BibitemOpen
	\bibfield  {author} {\bibinfo {author} {\bibfnamefont {W.}~\bibnamefont
			{Chen}}, \bibinfo {author} {\bibfnamefont {M.}~\bibnamefont {Abbasi}},
		\bibinfo {author} {\bibfnamefont {Y.~N.}\ \bibnamefont {Joglekar}},\ and\
		\bibinfo {author} {\bibfnamefont {K.~W.}\ \bibnamefont {Murch}},\ }\bibfield
	{title} {\bibinfo {title} {Quantum jumps in the non-{Hermitian} dynamics of a
			superconducting qubit},\ }\href
	{https://doi.org/10.1103/PhysRevLett.127.140504} {\bibfield  {journal}
		{\bibinfo  {journal} {Phys. Rev. Lett.}\ }\textbf {\bibinfo {volume} {127}},\
		\bibinfo {pages} {140504} (\bibinfo {year} {2021})}\BibitemShut {NoStop}%
	\bibitem [{\citenamefont {Han}\ \emph {et~al.}(2025)\citenamefont {Han},
		\citenamefont {Wu}, \citenamefont {Huang}, \citenamefont {Wu}, \citenamefont
		{Zou}, \citenamefont {Yi}, \citenamefont {Zhang}, \citenamefont {Li},
		\citenamefont {Xu}, \citenamefont {Zheng}, \citenamefont {Fan}, \citenamefont
		{Wen}, \citenamefont {Yang},\ and\ \citenamefont {Zheng}}]{Han2025p2400446}%
	\BibitemOpen
	\bibfield  {author} {\bibinfo {author} {\bibfnamefont {P.-R.}\ \bibnamefont
			{Han}}, \bibinfo {author} {\bibfnamefont {F.}~\bibnamefont {Wu}}, \bibinfo
		{author} {\bibfnamefont {X.-J.}\ \bibnamefont {Huang}}, \bibinfo {author}
		{\bibfnamefont {H.-Z.}\ \bibnamefont {Wu}}, \bibinfo {author} {\bibfnamefont
			{C.-L.}\ \bibnamefont {Zou}}, \bibinfo {author} {\bibfnamefont
			{W.}~\bibnamefont {Yi}}, \bibinfo {author} {\bibfnamefont {M.}~\bibnamefont
			{Zhang}}, \bibinfo {author} {\bibfnamefont {H.}~\bibnamefont {Li}}, \bibinfo
		{author} {\bibfnamefont {K.}~\bibnamefont {Xu}}, \bibinfo {author}
		{\bibfnamefont {D.}~\bibnamefont {Zheng}}, \bibinfo {author} {\bibfnamefont
			{H.}~\bibnamefont {Fan}}, \bibinfo {author} {\bibfnamefont {J.}~\bibnamefont
			{Wen}}, \bibinfo {author} {\bibfnamefont {Z.-B.}\ \bibnamefont {Yang}},\ and\
		\bibinfo {author} {\bibfnamefont {S.-B.}\ \bibnamefont {Zheng}},\ }\bibfield
	{title} {\bibinfo {title} {Enhancement of sensitivity near exceptional points
			in dissipative qubit--resonator systems},\ }\href
	{https://doi.org/10.1002/qute.202400446} {\bibfield  {journal} {\bibinfo
			{journal} {Adv. Quantum Technol.}\ }\textbf {\bibinfo {volume} {8}},\
		\bibinfo {pages} {2400446} (\bibinfo {year} {2025})}\BibitemShut {NoStop}%
	\bibitem [{\citenamefont {Wang}\ and\ \citenamefont
		{Zhou}(2014)}]{Wang2014p013809}%
	\BibitemOpen
	\bibfield  {author} {\bibinfo {author} {\bibfnamefont {Z.}~\bibnamefont
			{Wang}}\ and\ \bibinfo {author} {\bibfnamefont {D.~L.}\ \bibnamefont
			{Zhou}},\ }\bibfield  {title} {\bibinfo {title} {Quasidark state and quantum
			interference in the {Jaynes--Cummings} model with a common bath},\ }\href
	{https://doi.org/10.1103/PhysRevA.89.013809} {\bibfield  {journal} {\bibinfo
			{journal} {Phys. Rev. A}\ }\textbf {\bibinfo {volume} {89}},\ \bibinfo
		{pages} {013809} (\bibinfo {year} {2014})}\BibitemShut {NoStop}%
	\bibitem [{\citenamefont {Moore}(1920)}]{Moore1920p394}%
	\BibitemOpen
	\bibfield  {author} {\bibinfo {author} {\bibfnamefont {E.~H.}\ \bibnamefont
			{Moore}},\ }\bibfield  {title} {\bibinfo {title} {On the reciprocal of the
			general algebraic matrix},\ }\href@noop {} {\bibfield  {journal} {\bibinfo
			{journal} {Bull. Am. Math. Soc.}\ }\textbf {\bibinfo {volume} {26}},\
		\bibinfo {pages} {394} (\bibinfo {year} {1920})}\BibitemShut {NoStop}%
	\bibitem [{\citenamefont {Penrose}(1955)}]{Penrose1955p406}%
	\BibitemOpen
	\bibfield  {author} {\bibinfo {author} {\bibfnamefont {R.}~\bibnamefont
			{Penrose}},\ }\bibfield  {title} {\bibinfo {title} {A generalized inverse for
			matrices},\ }\href {https://doi.org/10.1017/S0305004100030401} {\bibfield
		{journal} {\bibinfo  {journal} {Proc. Cambridge Philos. Soc.}\ }\textbf
		{\bibinfo {volume} {51}},\ \bibinfo {pages} {406} (\bibinfo {year}
		{1955})}\BibitemShut {NoStop}%
	\bibitem [{\citenamefont {Baksalary}\ and\ \citenamefont
		{Trenkler}(2021)}]{Baksalary2021p9}%
	\BibitemOpen
	\bibfield  {author} {\bibinfo {author} {\bibfnamefont {O.~M.}\ \bibnamefont
			{Baksalary}}\ and\ \bibinfo {author} {\bibfnamefont {G.}~\bibnamefont
			{Trenkler}},\ }\bibfield  {title} {\bibinfo {title} {The {Moore--Penrose}
			inverse: a hundred years on a frontline of physics research},\ }\href
	{https://doi.org/10.1140/epjh/s13129-021-00011-y} {\bibfield  {journal}
		{\bibinfo  {journal} {Eur. Phys. J. H}\ }\textbf {\bibinfo {volume} {46}},\
		\bibinfo {pages} {9} (\bibinfo {year} {2021})}\BibitemShut {NoStop}%
	\bibitem [{\citenamefont {Rivas}\ and\ \citenamefont
		{Huelga}(2012)}]{RivasHuelga2012}%
	\BibitemOpen
	\bibfield  {author} {\bibinfo {author} {\bibfnamefont {A.}~\bibnamefont
			{Rivas}}\ and\ \bibinfo {author} {\bibfnamefont {S.~F.}\ \bibnamefont
			{Huelga}},\ }\href {https://doi.org/10.1007/978-3-642-23354-8} {\emph
		{\bibinfo {title} {Open Quantum Systems: An Introduction}}},\ SpringerBriefs
	in Physics\ (\bibinfo  {publisher} {Springer},\ \bibinfo {address} {Berlin,
		Heidelberg},\ \bibinfo {year} {2012})\BibitemShut {NoStop}%
	\bibitem [{\citenamefont {Breuer}\ and\ \citenamefont
		{Petruccione}(2007)}]{Breuer2007}%
	\BibitemOpen
	\bibfield  {author} {\bibinfo {author} {\bibfnamefont {H.-P.}\ \bibnamefont
			{Breuer}}\ and\ \bibinfo {author} {\bibfnamefont {F.}~\bibnamefont
			{Petruccione}},\ }\href
	{https://doi.org/10.1093/acprof:oso/9780199213900.001.0001} {\emph {\bibinfo
			{title} {The Theory of Open Quantum Systems}}}\ (\bibinfo  {publisher}
	{Oxford University Press},\ \bibinfo {address} {Oxford},\ \bibinfo {year}
	{2007})\BibitemShut {NoStop}%
	\bibitem [{\citenamefont {Trushechkin}(2021)}]{Trushechkin2021p062226}%
	\BibitemOpen
	\bibfield  {author} {\bibinfo {author} {\bibfnamefont {A.}~\bibnamefont
			{Trushechkin}},\ }\bibfield  {title} {\bibinfo {title} {Unified
			{Gorini--Kossakowski--Lindblad--Sudarshan} quantum master equation beyond the
			secular approximation},\ }\href {https://doi.org/10.1103/PhysRevA.103.062226}
	{\bibfield  {journal} {\bibinfo  {journal} {Phys. Rev. A}\ }\textbf {\bibinfo
			{volume} {103}},\ \bibinfo {pages} {062226} (\bibinfo {year}
		{2021})}\BibitemShut {NoStop}%
	\bibitem [{\citenamefont {Horn}\ and\ \citenamefont
		{Johnson}(2013)}]{HornJohnson2013}%
	\BibitemOpen
	\bibfield  {author} {\bibinfo {author} {\bibfnamefont {R.~A.}\ \bibnamefont
			{Horn}}\ and\ \bibinfo {author} {\bibfnamefont {C.~R.}\ \bibnamefont
			{Johnson}},\ }\href {https://doi.org/10.1017/CBO9781139020411} {\emph
		{\bibinfo {title} {Matrix Analysis}}},\ \bibinfo {edition} {2nd}\ ed.\
	(\bibinfo  {publisher} {Cambridge University Press},\ \bibinfo {address}
	{Cambridge},\ \bibinfo {year} {2013})\BibitemShut {NoStop}%
	\bibitem [{\citenamefont {Gorini}\ \emph {et~al.}(1976)\citenamefont {Gorini},
		\citenamefont {Kossakowski},\ and\ \citenamefont
		{Sudarshan}}]{Gorini1976p821}%
	\BibitemOpen
	\bibfield  {author} {\bibinfo {author} {\bibfnamefont {V.}~\bibnamefont
			{Gorini}}, \bibinfo {author} {\bibfnamefont {A.}~\bibnamefont
			{Kossakowski}},\ and\ \bibinfo {author} {\bibfnamefont {E.~C.~G.}\
			\bibnamefont {Sudarshan}},\ }\bibfield  {title} {\bibinfo {title} {Completely
			positive dynamical semigroups of {N}-level systems},\ }\href
	{https://doi.org/10.1063/1.522979} {\bibfield  {journal} {\bibinfo  {journal}
			{J. Math. Phys.}\ }\textbf {\bibinfo {volume} {17}},\ \bibinfo {pages} {821}
		(\bibinfo {year} {1976})}\BibitemShut {NoStop}%
	\bibitem [{\citenamefont {Lindblad}(1976)}]{Lindblad1976p119}%
	\BibitemOpen
	\bibfield  {author} {\bibinfo {author} {\bibfnamefont {G.}~\bibnamefont
			{Lindblad}},\ }\bibfield  {title} {\bibinfo {title} {On the generators of
			quantum dynamical semigroups},\ }\href {https://doi.org/10.1007/BF01608499}
	{\bibfield  {journal} {\bibinfo  {journal} {Commun. Math. Phys.}\ }\textbf
		{\bibinfo {volume} {48}},\ \bibinfo {pages} {119} (\bibinfo {year}
		{1976})}\BibitemShut {NoStop}%
	\bibitem [{\citenamefont {Cattaneo}\ \emph {et~al.}(2019)\citenamefont
		{Cattaneo}, \citenamefont {Giorgi}, \citenamefont {Maniscalco},\ and\
		\citenamefont {Zambrini}}]{Cattaneo2019p113045}%
	\BibitemOpen
	\bibfield  {author} {\bibinfo {author} {\bibfnamefont {M.}~\bibnamefont
			{Cattaneo}}, \bibinfo {author} {\bibfnamefont {G.~L.}\ \bibnamefont
			{Giorgi}}, \bibinfo {author} {\bibfnamefont {S.}~\bibnamefont {Maniscalco}},\
		and\ \bibinfo {author} {\bibfnamefont {R.}~\bibnamefont {Zambrini}},\
	}\bibfield  {title} {\bibinfo {title} {Local versus global master equation
			with common and separate baths: Superiority of the global approach in partial
			secular approximation},\ }\href {https://doi.org/10.1088/1367-2630/ab54ac}
	{\bibfield  {journal} {\bibinfo  {journal} {New J. Phys.}\ }\textbf {\bibinfo
			{volume} {21}},\ \bibinfo {pages} {113045} (\bibinfo {year}
		{2019})}\BibitemShut {NoStop}%
	\bibitem [{\citenamefont {Kafuri}\ \emph {et~al.}(2024)\citenamefont {Kafuri},
		\citenamefont {Maldonado-Villamizar}, \citenamefont {Moroz},\ and\
		\citenamefont {Rodr{\'i}guez-Lara}}]{Kafuri2024pC82}%
	\BibitemOpen
	\bibfield  {author} {\bibinfo {author} {\bibfnamefont {A.}~\bibnamefont
			{Kafuri}}, \bibinfo {author} {\bibfnamefont {F.~H.}\ \bibnamefont
			{Maldonado-Villamizar}}, \bibinfo {author} {\bibfnamefont {A.}~\bibnamefont
			{Moroz}},\ and\ \bibinfo {author} {\bibfnamefont {B.~M.}\ \bibnamefont
			{Rodr{\'i}guez-Lara}},\ }\bibfield  {title} {\bibinfo {title} {Supersymmetry
			journey from the {Jaynes--Cummings} to the anisotropic {Rabi} model},\ }\href
	{https://doi.org/10.1364/JOSAB.522504} {\bibfield  {journal} {\bibinfo
			{journal} {J. Opt. Soc. Am. B}\ }\textbf {\bibinfo {volume} {41}},\ \bibinfo
		{pages} {C82} (\bibinfo {year} {2024})}\BibitemShut {NoStop}%
	\bibitem [{\citenamefont {van Loo}\ \emph {et~al.}(2013)\citenamefont {van
			Loo}, \citenamefont {Fedorov}, \citenamefont {Lalumi{\`e}re}, \citenamefont
		{Sanders}, \citenamefont {Blais},\ and\ \citenamefont
		{Wallraff}}]{vanLoo2013p1494}%
	\BibitemOpen
	\bibfield  {author} {\bibinfo {author} {\bibfnamefont {A.~F.}\ \bibnamefont
			{van Loo}}, \bibinfo {author} {\bibfnamefont {A.}~\bibnamefont {Fedorov}},
		\bibinfo {author} {\bibfnamefont {K.}~\bibnamefont {Lalumi{\`e}re}}, \bibinfo
		{author} {\bibfnamefont {B.~C.}\ \bibnamefont {Sanders}}, \bibinfo {author}
		{\bibfnamefont {A.}~\bibnamefont {Blais}},\ and\ \bibinfo {author}
		{\bibfnamefont {A.}~\bibnamefont {Wallraff}},\ }\bibfield  {title} {\bibinfo
		{title} {Photon-mediated interactions between distant artificial atoms},\
	}\href {https://doi.org/10.1126/science.1244324} {\bibfield  {journal}
		{\bibinfo  {journal} {Science}\ }\textbf {\bibinfo {volume} {342}},\ \bibinfo
		{pages} {1494} (\bibinfo {year} {2013})}\BibitemShut {NoStop}%
	\bibitem [{\citenamefont {Wang}\ \emph {et~al.}(2019)\citenamefont {Wang},
		\citenamefont {Du}, \citenamefont {Li},\ and\ \citenamefont
		{x.~Liu}}]{Wang2019p053809}%
	\BibitemOpen
	\bibfield  {author} {\bibinfo {author} {\bibfnamefont {Z.}~\bibnamefont
			{Wang}}, \bibinfo {author} {\bibfnamefont {L.}~\bibnamefont {Du}}, \bibinfo
		{author} {\bibfnamefont {Y.}~\bibnamefont {Li}},\ and\ \bibinfo {author}
		{\bibfnamefont {Y.}~\bibnamefont {x.~Liu}},\ }\bibfield  {title} {\bibinfo
		{title} {Phase-controlled single-photon nonreciprocal transmission in a
			one-dimensional waveguide},\ }\href
	{https://doi.org/10.1103/PhysRevA.100.053809} {\bibfield  {journal} {\bibinfo
			{journal} {Phys. Rev. A}\ }\textbf {\bibinfo {volume} {100}},\ \bibinfo
		{pages} {053809} (\bibinfo {year} {2019})}\BibitemShut {NoStop}%
	\bibitem [{\citenamefont {Meekhof}\ \emph {et~al.}(1996)\citenamefont
		{Meekhof}, \citenamefont {Monroe}, \citenamefont {King}, \citenamefont
		{Itano},\ and\ \citenamefont {Wineland}}]{Meekhof1996p1796}%
	\BibitemOpen
	\bibfield  {author} {\bibinfo {author} {\bibfnamefont {D.~M.}\ \bibnamefont
			{Meekhof}}, \bibinfo {author} {\bibfnamefont {C.}~\bibnamefont {Monroe}},
		\bibinfo {author} {\bibfnamefont {B.~E.}\ \bibnamefont {King}}, \bibinfo
		{author} {\bibfnamefont {W.~M.}\ \bibnamefont {Itano}},\ and\ \bibinfo
		{author} {\bibfnamefont {D.~J.}\ \bibnamefont {Wineland}},\ }\bibfield
	{title} {\bibinfo {title} {Generation of nonclassical motional states of a
			trapped atom},\ }\href {https://doi.org/10.1103/PhysRevLett.76.1796}
	{\bibfield  {journal} {\bibinfo  {journal} {Phys. Rev. Lett.}\ }\textbf
		{\bibinfo {volume} {76}},\ \bibinfo {pages} {1796} (\bibinfo {year}
		{1996})}\BibitemShut {NoStop}%
	\bibitem [{\citenamefont {So}\ \emph {et~al.}(2024)\citenamefont {So},
		\citenamefont {Suganthi}, \citenamefont {Menon}, \citenamefont {Zhu},
		\citenamefont {Zhuravel}, \citenamefont {Pu}, \citenamefont {Wolynes},
		\citenamefont {Onuchic},\ and\ \citenamefont {Pagano}}]{So2024peads8011}%
	\BibitemOpen
	\bibfield  {author} {\bibinfo {author} {\bibfnamefont {V.}~\bibnamefont
			{So}}, \bibinfo {author} {\bibfnamefont {M.~D.}\ \bibnamefont {Suganthi}},
		\bibinfo {author} {\bibfnamefont {A.}~\bibnamefont {Menon}}, \bibinfo
		{author} {\bibfnamefont {M.}~\bibnamefont {Zhu}}, \bibinfo {author}
		{\bibfnamefont {R.}~\bibnamefont {Zhuravel}}, \bibinfo {author}
		{\bibfnamefont {H.}~\bibnamefont {Pu}}, \bibinfo {author} {\bibfnamefont
			{P.~G.}\ \bibnamefont {Wolynes}}, \bibinfo {author} {\bibfnamefont {J.~N.}\
			\bibnamefont {Onuchic}},\ and\ \bibinfo {author} {\bibfnamefont
			{G.}~\bibnamefont {Pagano}},\ }\bibfield  {title} {\bibinfo {title}
		{Trapped-ion quantum simulation of electron transfer models with tunable
			dissipation},\ }\href {https://doi.org/10.1126/sciadv.ads8011} {\bibfield
		{journal} {\bibinfo  {journal} {Sci. Adv.}\ }\textbf {\bibinfo {volume}
			{10}},\ \bibinfo {pages} {eads8011} (\bibinfo {year} {2024})}\BibitemShut
	{NoStop}%
	\bibitem [{\citenamefont {Junge}\ \emph {et~al.}(2013)\citenamefont {Junge},
		\citenamefont {O'Shea}, \citenamefont {Volz},\ and\ \citenamefont
		{Rauschenbeutel}}]{Junge2013p213604}%
	\BibitemOpen
	\bibfield  {author} {\bibinfo {author} {\bibfnamefont {C.}~\bibnamefont
			{Junge}}, \bibinfo {author} {\bibfnamefont {D.}~\bibnamefont {O'Shea}},
		\bibinfo {author} {\bibfnamefont {J.}~\bibnamefont {Volz}},\ and\ \bibinfo
		{author} {\bibfnamefont {A.}~\bibnamefont {Rauschenbeutel}},\ }\bibfield
	{title} {\bibinfo {title} {Strong coupling between single atoms and
			nontransversal photons},\ }\href
	{https://doi.org/10.1103/PhysRevLett.110.213604} {\bibfield  {journal}
		{\bibinfo  {journal} {Phys. Rev. Lett.}\ }\textbf {\bibinfo {volume} {110}},\
		\bibinfo {pages} {213604} (\bibinfo {year} {2013})}\BibitemShut {NoStop}%
	\bibitem [{\citenamefont {Scheucher}\ \emph {et~al.}(2016)\citenamefont
		{Scheucher}, \citenamefont {Hilico}, \citenamefont {Will}, \citenamefont
		{Volz},\ and\ \citenamefont {Rauschenbeutel}}]{Scheucher2016p1577}%
	\BibitemOpen
	\bibfield  {author} {\bibinfo {author} {\bibfnamefont {M.}~\bibnamefont
			{Scheucher}}, \bibinfo {author} {\bibfnamefont {A.}~\bibnamefont {Hilico}},
		\bibinfo {author} {\bibfnamefont {E.}~\bibnamefont {Will}}, \bibinfo {author}
		{\bibfnamefont {J.}~\bibnamefont {Volz}},\ and\ \bibinfo {author}
		{\bibfnamefont {A.}~\bibnamefont {Rauschenbeutel}},\ }\bibfield  {title}
	{\bibinfo {title} {Quantum optical circulator controlled by a single chirally
			coupled atom},\ }\href {https://doi.org/10.1126/science.aaj2118} {\bibfield
		{journal} {\bibinfo  {journal} {Science}\ }\textbf {\bibinfo {volume}
			{354}},\ \bibinfo {pages} {1577} (\bibinfo {year} {2016})}\BibitemShut
	{NoStop}%
\end{thebibliography}
%

\end{document}